\documentclass[runningheads]{llncs}
\usepackage[english]{babel}
\usepackage[utf8]{inputenc}
\usepackage{amssymb}

\spnewtheorem{myclaim}[Piotrekclm]{Claim}{\bfseries}{\itshape}
\usepackage{thm-restate}

\usepackage{mathtools}

\usepackage{graphicx}
\usepackage[linesnumbered,ruled,vlined]{algorithm2e}
\usepackage{todonotes}

\newcommand{\orange}[1]{\textcolor{orange}{#1}}

\newcommand{\removed}[1]{}

\usepackage[utf8]{inputenc}
\usepackage{glossaries}

\newcommand{\da}{\alpha}

\newcommand{\uniformize}{\mathit{uniformize}}
\newcommand{\replace}{\mathit{replace}}
\newcommand{\rotate}{\mathit{rotate}}
\newcommand{\concat}{\concatenate}
\newcommand{\decrease}{\mathit{decrease}}
\newcommand{\uniformyzing}{\mathcal{U}}
\newcommand{\reducing}{\mathcal{R}}
\newcommand{\rotating}{\mathit{rot}}
\newcommand{\decreasing}{\mathit{dec}}

\newcommand{\Q}{\mathbb{Q}}
\newcommand{\N}{\mathbb{N}}
\newcommand{\Z}{\mathbb{Z}}
\newcommand{\X}{\mathbb{X}}
\newcommand{\D}{\mathbb{D}}
\newcommand{\marki}{{\textbf{\textit{i}}}}
\newcommand{\markf}{{\textbf{\textit{f}}}}
\newcommand{\markm}{\textbf{\textit{m}}}
\newcommand{\markz}{\textbf{\textit{z}}}
\newcommand{\run}[2]{\{(#1)\}_{#2}}

\newcommand{\Var}{\mathit{Var}}
\newcommand{\dval}{\mathit{dval}}
\newcommand{\var}{\mathit{vars}}

\newcommand{\row}{\mathit{row}}
\newcommand{\col}{\mathit{col}}

\newcommand{\Hist}{Hist}

\newcommand{\db}{\beta}

\newcommand{\setS}{\mathbb{S}}
\newcommand{\setE}{{\mathbb{E}}}
\newcommand{\concatenate}{\copyright}

\newcommand{\Pmat}{{\mathcal{P}}}
\newcommand{\cN}{\mathcal{N}}
\newcommand{\setY}{\mathbb{Y}}
\newcommand{\combined}{\bigotimes}
\newcommand{\Fbul}{\orange{F}}

\newcommand{\PTIME}{PTime}

\newcommand{\ETIME}{ExpTime}

\newcommand{\Net}[1]{(#1)}

\newcommand{\ESPACE}{ExpSpace}

\newcommand{\loopless}{loop-less}
\newcommand{\zero}{\textbf{0}}

\usepackage{tikz}
\usepackage{blkarray}
\usetikzlibrary{arrows,shapes,automata,backgrounds,petri}
\usetikzlibrary{positioning,chains,fit,shapes,calc}

\newcommand{\repeatthanks}{\textsuperscript{\thefootnote}}

\title{Continuous Reachability for Unordered Data Petri nets is in PTime\thanks{Supported by Polish NCN grant UMO-2016/21/D/ST6/01368, DST Inspire faculty award IFA12-MA-17.}}

\authorrunning{Utkarsh Gupta, Preey Shah, S. Akshay and Piotr Hofman}

\author{Utkarsh Gupta\thanks{Both these authors contributed equally to this work.}\inst{1} \and Preey Shah\repeatthanks\inst{1} \and S. Akshay\inst{1} \and Piotr Hofman\inst{2}}

\institute{Department of CSE, IIT Bombay, India
\and 
University of Warsaw, Poland}

\begin{document}

\maketitle

\begin{abstract}

Unordered data Petri nets (UDPN) are an extension of classical Petri nets with tokens that carry data from an infinite domain and where transitions may check equality and disequality of tokens. UDPN are well-structured, so the coverability and termination problems are decidable, but with higher complexity than for Petri nets. On the other hand, the problem of reachability for UDPN is surprisingly complex, and its decidability status remains open.
In this paper, we consider the continuous reachability problem for UDPN, which can be seen as an over-approximation of the reachability problem. Our main result is a characterization of continuous reachability for UDPN and polynomial time algorithm for solving it.   This is a consequence of a combinatorial argument, which shows that if continuous reachability holds then there exists a run using only polynomially many data values.

\end{abstract}

\section{Introduction}
The theory of Petri Nets has been developing since more than 50 years. On one hand, from a theory perspective, Petri Nets are interesting due to their deep mathematical structure and despite exhibiting nice properties, like being a well structured transition system \cite{WSTS-everywhere}, we still don't understand them well. On the other hand, Petri Nets are a useful pictorial formalism for modeling and thus found their way to the industry. To connect this theory and practice, it would be desirable to use the developed theory of Petri Nets \cite{DBLP:journals/tcs/Rackoff78,DBLP:conf/stoc/Kosaraju82,DBLP:conf/lics/LerouxS15} for the symbolic analysis and verification of Petri Nets models. However, we already know that this is difficult in its full generality. It suffices to recall two results that were proved more than 30 years apart. An old but classical result by Lipton~\cite{Lipton} shows that even coverability is \ESPACE-hard, while the non-elementary hardness of the reachability relation has just been established this year \cite{Czerwinski}.
Moreover, when we look at Petri nets based formalisms that are needed to model various aspects of industrial systems, we see that they go beyond the expressivity of Petri Nets. For instance, colored Petri nets, which are used in modeling workflows~\cite{DBLP:journals/jcsc/Aalst98}, allow the tokens to be colored with an infinite set of colors, and introduce a complex formalism to describe dependencies between colors. This makes all verification problems undecidable for this generic model.
Given the basic nature and importance of the reachability problem in Petri nets (and its extensions), there have been several efforts to sidestep the complexity-theoretic hardness results. One common approach is to look for easy subclasses (such as bounded nets~\cite{DBLP:conf/ac/Esparza96}, free-choice nets~\cite{Desel:1995:FCP:207572} etc). The other approach, which we adopt in this work, is to compute over-approximations of the reachability relation. 

{\it Continuous reachability.} A natural question regarding the dynamics of a Petri net is to ask what would happen if tokens instead of behaving like  discrete units start to behave like a continuous fluid? This simple question led us to an elegant theory of so-called continuous Petri nets \cite{Continuous-original,DBLP:journals/fuin/FracaH15,DBLP:conf/lics/BlondinH17}. Petri nets with continuous semantics allow markings to be functions from places to \emph{nonnegative rational numbers} (i.e., in $\Q^+$) instead of natural numbers. Moreover, whenever a transition is fired a positive rational coefficient is chosen and both the number of consumed and produced tokens are multiplied with the coefficient.
This allows to split tokens into arbitrarily small parts and process them independently. 
This for instance may occur in applications related to hybrid systems where the discrete part is used to control the continuous systems \cite{DBLP:journals/automatica/DavidA94,Alla1998ContinuousAH}.
Interestingly, this makes things simpler to analyze. 
For example reachability under the continuous semantics for Petri nets is $\PTIME$-complete \cite{DBLP:journals/fuin/FracaH15}. 

However, when one wants to analyze extensions of Petri nets, for example reset Petri Nets with continuous semantics, it turns out that reachability is as hard as reachability in reset Petri nets under the usual semantics i.e. it is undecidable\footnote{This can be seen on the same lines as the proof of undecidability of continuous reachability for Petri nets with zero tests~\cite{DBLP:conf/lics/BlondinH17}.}. In this paper we identify an extension of Petri nets with unordered data, for which this is not the case and continuous semantics leads to a substantial reduction in the complexity of the reachability problem.

{\it Unordered data Petri Nets.}
The possibility of equipping tokens with some additional information is one of the main lines of research regarding extensions of Petri Nets, the best known being Colored Petri Nets
\cite{ColoredPetriNets} and various types of timed Petri Nets \cite{Wang1998,Abdulla:2001:TPN:647747.734218}. 
In \cite{Nets-with-Tokens-which-Carry-Data} authors equipped tokens with data and restricted interactions between data in a way that allow to transfer techniques for well structured transition systems.
They identified various classes of nets exhibiting interesting combinatorial properties which led to a number of results
\cite{Forward-Analysis-for-Petri-Nets-with-Name-Creation,What-Makes-Petri-Nets-Harder-to-Verify-Stack-or-Data,Coverability-Trees-for-Petri-Nets-with-Unordered-Data,DBLP:conf/lics/HofmanLT17,DBLP:conf/concur/HofmanL18}.
    Unordered Data Petri Nets (UDPN), are simplest among them: every token carries a single datum like a barcode and transitions may check equality or disequality of data in consumed and produced tokens. UDPN are the only class identified in \cite{Nets-with-Tokens-which-Carry-Data} for which the reachability is still unsolved, although in~\cite{What-Makes-Petri-Nets-Harder-to-Verify-Stack-or-Data} authors show that the problem is at least Ackermannian-hard (for all other data extensions, reachability is undecidable). A recent attempt to over-approximate the reachability relation for UDPN in~\cite{DBLP:conf/lics/HofmanLT17} considers integer reachability i.e. number of tokens may get negative during the run (also called solution of the state equation). From the above perspective, this paper is an extension of the mentioned line of research. 

{\it Our contribution.} Our main contribution is a characterization of continuous reachability in UDPN and a polynomial time algorithm for solving it.
Observe that if we find an upper bound on the minimal number of data required by a run between two configurations (if any run exists), then we can reduce continuous reachability in UDPN to continuous reachability in vanilla Petri nets with an exponential blowup and use the already developed characterization from \cite{DBLP:journals/fuin/FracaH15}.
In Section~\ref{sec:Bound} we prove such a bound on the minimal number of required data. The bound is novel and exploits techniques that did not appear previously in the context of data nets. Further, the obtained bounds are lower than bounds on the number of data values required to solve the state equation \cite{DBLP:conf/lics/HofmanLT17}, which is surprising considering that existence of a continuous run requires a solution of a sort of state equation. Precisely, the difference is that we are looking for solutions of the state equation over $\Q^+$ instead of $\N$ and in this case we prove better bounds for the number of data required. This also gives us an easy polytime algorithm for finding $\Q^+$-solutions of state equations of UDPN 
(we remark that for Petri nets without data, this appears among standard algebraic techniques ~\cite{Silva1998}).

Finally, with the above bound, we solve continuous reachability in UDPN by adapting the techniques from the non-data setting of \cite{DBLP:conf/lics/BlondinH17,DBLP:conf/tacas/BlondinFHH16}. 
We adapt the characterization of continuous reachability to the data setting and next encode it as system of linear equations with implications. In doing so, however, we face the problem that a naive encoding (representing data explicitly) gives a system of equations of exponential size, giving only an \ETIME-algorithm. To improve the complexity, we use histograms, a combinatorial tool developed in \cite{DBLP:conf/lics/HofmanLT17}, to compress the description of solutions of state equations in UDPNs. However, this may lead to spurious solutions for continuous reachability. To eliminate them, we show that it suffices to first transform the net and then apply the idea of histograms to characterize continuous runs in the modified net. The whole procedure is described in
Section~\ref{Sec:Q+algorithm} and leads us to our $\PTIME$ algorithm for continuous reachability in UDPN. Note that since we easily have $\PTIME$ hardness for the problem (even without data), we obtain that the problem of continuous reachability in UDPN is $\PTIME$-complete.

{\it Towards verification.} Over-approximations are useful in verification of Petri nets and their extensions: as explained in~\cite{Silva1998}, for many practical problems, over-approximate solutions are already correct. Further, we can use them as a sub-routine to improve the practical performance of verification algorithms. A remarkable example is the recent work in~\cite{DBLP:conf/tacas/BlondinFHH16}, where the $\PTIME$ continuous reachability algorithm for Petri nets from~\cite{DBLP:journals/fuin/FracaH15} is used as a subroutine to solve the $\ESPACE$ hard coverability problem in Petri nets, outperforming the best known tools for this problem, such as Petrinizer~\cite{Petrinizer}. Our results can be seen as a first step in the same spirit towards handling practical instances of coverability, but for the extended model of UDPN, 
where the coverability problem for UDPN is known to be Ackermannian-hard~\cite{What-Makes-Petri-Nets-Harder-to-Verify-Stack-or-Data}.

\section{Preliminaries} \label{sec:Prelim}
We denote integers, non-negative integers, rationals, and reals as $\Z,\N,\mathbb{Q},$ and $\mathbb{R}$, respectively. For a set $\X \subseteq \mathbb{R}$ denote by $\X^{+}$, the set of all non-negative elements of $\X$. 
 We denote by \textbf{0}, a vector whose entries are all zero. We define in a standard point-wise way operations on vectors i.e. \emph{scalar multiplication $\cdot$, addition $+$, subtraction $-$, and vector comparison $\leq$.}
 In this paper, we use functions of the type $X\to (Y\to Z)$, and instead of $(f(x))(y)$, we write $f(y,x).$ 
For functions $f,g$ where the range of $g$ is a subset of the domain of $f$, we denote their composition by $f \circ g$. If $\pi$ is an injection then by $\pi^{-1}$ we mean a partial function such that $\pi^{-1}\circ \pi$ is the identity function.
Let $f : X_1 \to Y$ , $g: X_2 \to Y$ be two functions with addition and scalar multiplication operations defined on $Y.$ 
A \emph{scalar multiplication} of a function is defined as follows $(a\cdot f)(x)=a\cdot f(x)$ for all $x\in X_1.$
We lift \emph{addition} operation to functions pointwise, i.e. $f + g : X_1\cup X_2 \to Y$ such that 
 \begin{equation*}
     (f+g)(x)=\begin{cases*}
      f(x) & if $x\in X_1\setminus X_2$ \\
      g(x) & if $x\in X_2\setminus X_1$ \\
      f(x)+g(x) & if $x\in X_1\cap X_2$.
    \end{cases*}
 \end{equation*} Similarly for \emph{subtraction} $(f-g)(x) = f(x)+-1\cdot g(x)$.

We use \emph{matrices} with \emph{rows and columns} indexed by sets $\setS_1,\setS_2$, possibly infinite. For a matrix $M$, let $M(r,c)$ denote the entry at column $c$ and row $r$, and $M(r,\bullet)$,$M(\bullet,c)$ denote the row vector indexed by $r$ and column vector indexed by $c$, respectively. Denote by $\col(M)$, $\row(M)$ the set of indices of nonzero columns and nonzero rows of the matrix $M$, respectively. Even if we have infinitely many rows or columns, our matrices will have only finitely many \emph{nonzero} rows and columns, and only this nonzero part will be represented. Following our nonstandard matrix definition we precisely define operations on them, although they are natural. First, a \emph{multiplication by a constant number} produces a new matrix with row and columns labelled with the same sets $\setS_1, \setS_2$ and defined as follows $(a\cdot M)(r,c)=a\cdot(M(r,c))$ for all  $(r,c)\in \setS_1\times \setS_2$. \emph{Addition} of two matrices is only defined if the sets indexing rows $\setS_1$ and columns $\setS_2$ are the same for both summands $M_1$ and $M_2$,  $\forall(r,c)\in \setS_1\times\setS_2$ the sum $(M_1+M_2)(r,c)=M_1(r,c)+M_2(r,c)$, the \emph{subtraction} $M_1-M_2$ is a shorthand for $M_1+ (-1)\cdot M_2$. Observe that all but finitely many entries in matrices are $0$, and therefore when we do computation on matrices we can restrict to rows $\row(M_1)\cup\row(M_2)$ and columns $\col(M_1)\cup \col(M_2)$. Similarly the \emph{comparison} 
for two matrices $M_1,M_2$ is defined as follows $M_1 \leq M_2$  if $\forall (r,c)\in (\row(M_1)\cup \row(M_2))\times (\col(M_1)\cup \col(M_2)) ~ M_1(r,c) \leq M_2(r,c)$; relations $>,\geq,\leq$ are defined analogically. The last operation which we need is matrix multiplication $M_1\cdot M_2=M_3$, it is only allowed if the set of columns of the first matrix $M_1$ is the same as  the set of rows of the second matrix $M_2$, the sets of rows and columns of the resulting matrix $M_3$ are rows of the matrix $M_1$ and columns of $M_2$, respectively. $M_3(r,c)=\sum_{k}M_1(r,k)M_2(k,c)$ where $k$ runs through columns of $M_1.$ Again, observe that if the row or a column is equal to $0$ for all entries then the effect of multiplication is $0$, thus we may restrict to $\row(M_1)$ and $\col(M_2)$. Moreover in the sum it suffices to write $\sum_{k\in\col(M_1)}M_1(r,k)M_2(k,c).$

\section{UDPN, reachability and its variants: Our main results}

Unordered data Petri nets extend the classical model of Petri nets by allowing each token to hold a data value from a countably-infinite domain $\D$. Our definition is closest to the definition of $\nu$-Petri Nets from \cite{DBLP:journals/tcs/Rosa-VelardoF11}. 
For simplicity 
we choose this one instead of using the equivalent but complex one from \cite{Nets-with-Tokens-which-Carry-Data}.

\begin{definition} Let $\D$ be a countably infinite set. An unordered data Petri net (UDPN) over domain $\D$ is a tuple $\Net{P,T,F,\Var}$ where $P$ is a finite set of \textit{places}, $T$ is a finite set of \textit{transitions}, $\Var$ is a finite set of variables, and $F:(P \times T) \cup (T \times P) \to (\Var \to \N)$ is a flow function that assigns each place $p \in P$ and transition $t \in T$ a function over \textit{variables} in \textit{Var}.  
\end{definition}

For each transition $t \in T$ we define functions $\Fbul(\bullet,t)$ and $\Fbul(t,\bullet)$,  $\Var \to (P \to \N)$ as $\Fbul(\bullet,t)(p,x) = F(p,t)(x)$ and analogously $\Fbul(t,\bullet)(p,x) = F(t,p)(x)$. \emph{Displacement} of the transition $t$ is a function $\Delta(t):  \Var \to (P \to \Z)$ defined as $\Delta(t) \overset{\mathrm{def}}{=} F(t,\bullet)- F(\bullet,t)$.

For $\X\in \{\N, \Z,\Q,\Q^+\}$, we define an \emph{$\X$-marking} as a function $M: \D \to (P \to \X)$ that is constant $0$ on all except finitely many values of $\D$. Intuitively, $M(p, \da)$ denotes the number of \emph{tokens} with the data value $\da$ at place $p$.
The fact that it is $0$ at all but finitely many data means that the number of tokens in any $\X$-marking is finite. We denote the infinite set of all $\X$-markings by $\mathcal{M}_{\X}$.

 We define an \emph{$\X$-step} as a triple $(c,t,\pi)$ for a transition $t \in T$, \emph{mode} $\pi$ being an injective map $\pi: \Var \to \D$, and a scalar constant $c \in \X^{+}$.
 An $\X$-step $(c,t,\pi)$ is \emph{fireable} at a $\X$-marking $\marki$ if $\marki - c \cdot F(\bullet,t) \circ \pi^{-1}\in \mathcal{M}_{\X}.$ 
 
 The $\X$-marking $\markf$ reached after \emph{firing} an $\X$-step $(c,t,\pi)$ at $\marki$ is given as 
 $\markf = \marki + c \cdot \Delta(t) \circ \pi^{-1}$.
  We also say that 
  an $\X$-step $(c,t,\pi)$ when fired 
  \emph{consumes} tokens $c\cdot F(\bullet,t)\circ\pi^{-1}$ and
  \emph{produces} tokens $c\cdot F(t,\bullet)\circ\pi^{-1}$.
  We define an $\X$-run as a sequence of $\X$-steps and we can represent it as $\run{c_i,t_i,\pi_i}{|\rho|}$ where $(c_i,t_i,\pi_i)$ 
   is the $i^{th}$ $\X$-step and $|\rho|$ is the number of $\X$-steps.
   A run $\rho = \run{c_i,t_i,\pi_i}{|\rho|}$ is fireable at a $\X$-marking $\marki$ if, 
   $\forall 1 \leq i \leq |\rho|$, the step $(c_i, t_i,\pi_i)$ is fireable at $\marki + \sum_{j=1}^{i-1}c_i \Delta(t_j) \circ \pi_j^{-1}$. 
   By $\marki \xrightarrow[]{\rho}_{\X} \markf$ we denote that $\rho$ is fireable at $\marki$ and after firing $\rho$ at $\marki$ we reach $\X$-marking $\markf = \marki + \sum_{i=1}^{|\rho|} c_i \cdot \Delta(t_i)\circ \pi_i^{-1}$. We call (the  function computed by) the mentioned sum  $\sum_{i=1}^{|\rho|}c_i\Delta(t_i)\circ \pi_i^{-1}$
as the \emph{effect} of the run and denote it by $\Delta(\rho)$.

We fix some notations for the rest of the paper. We use Greek letters $\alpha,\beta,\gamma$ to denote data values from data domain $\D$, $\rho$, $\sigma$ to denote a run,  $\pi$ to denote a mode and $x,y,z$ to denote the variables. When clear from the context, we may omit $\X$ from $\X$-marking, $\X$-run and just write marking, run, etc. Further, we will use letters in bold, e.g., $\markm$ to denote markings, where $\marki,\markf$ will be used for initial and final markings respectively. Further, throughout the paper, unless stated explicitly otherwise, we will refer to a UDPN $\mathcal{N} = \Net{P,T,F, \Var}$, therefore $P,T,F, \Var$ will denote the places, transitions, flow, and variables of this UDPN. 

\begin{figure}
\centering
\begin{tikzpicture}[node distance=1.3cm,>=stealth',bend angle=45,auto]
  \tikzstyle{place}=[circle,thick,draw=blue!75,fill=blue!20,minimum size=6mm]
  \tikzstyle{red place}=[place,draw=red!75,fill=red!20]
  \tikzstyle{transition}=[rectangle,thick,draw=black!75,
  			  fill=black!20,minimum size=4mm]
  \begin{scope}
    \node [place,colored tokens={blue},label=$p_2$] (w1)[xshift=0mm,yshift=0mm]{};
    \node [transition] (t1) [left of=w1,xshift=0mm,yshift=-10mm] {t}
    edge [pre] node{x} (w1);
    \node [place,colored tokens={red,green},label=$p_1$] (w2)[left of=t1,xshift=0mm,yshift=10mm]{}
    edge [post] node{y}(t1);
    \node [place,colored tokens={red,blue},label=$p_4$] (p3)[below of=w1,xshift=0mm,yshift=-10mm]{}
    edge [pre] node{x,z}(t1);
    \node [place,colored tokens={red,red},label=$p_3$] (p4)[below of=w2,xshift=0mm,yshift=-10mm]{}
    edge [pre] node{$\{2y\}$}(t1);
  \end{scope}
\end{tikzpicture}
\caption{A simple UDPN $\cN_1$}
\label{fig:simpleUDPN}
\end{figure}
\begin{example}\label{ex:1}
An example of a simple UDPN $\cN_1$  is given in Figure~\ref{fig:simpleUDPN}. For this example, we have  $P=\{p_1,p_2,p_3,p_4\}$, $T=\{t\}$, $Var=\{x,y,z\}$ and the flow relation is given by  $F(p_1,t)=\{y\mapsto 1\}$, $F(p_2,t)=\{x\mapsto 1\}$, $F(t,p_3)=\{y\mapsto 2\}$, $F(t,p_4)=\{x\mapsto 1,z\mapsto 1\}$, and an assignment of $0$ to every variable for the remaining of the pairs. Thus, for enabling transition $p_1$ and $p_2$ must have one token each with a different data value (since $x\neq y$) and after firing two tokens are produced in $p_3$ with same data value as was consumed from $p_1$ and two tokens are produced in $p_4$, one of whom has same data as consumed from $p_2$.
\end{example}
 
\begin{definition}
Given $\X$-markings $\marki,\markf$, we say $\markf$ is  $\X$-reachable from $\marki$ if there exists an $\X$-run $\rho$ s.t., $\marki \xrightarrow{\rho}_{\X} \markf$.
\end{definition}

When $\X = \N$, $\X$-reachability is the classical reachability problem, whose decidability is still unknown, while $\Z$-reachability for UDPN is in NP~\cite{DBLP:conf/lics/HofmanLT17}. 
\removed{In a recent paper, it was shown that $\Z$-reachability for UDPN is decidable in NP~\cite{DBLP:conf/lics/HofmanLT17}. 
This is sometimes also phrased as saying that the state equation can be solved for UDPN in NP. The main idea was to bound the number data values needed in some  $\Z$-witness, and use this bound to reformulate the question as an integer linear program, which can be solved in NP. }
In this paper we tackle $\Q$ and $\Q^+$-reachability, also called \emph{continuous} reachability in UDPN. 

The first step towards the solution is 
showing that if a $\Q^{+}$-marking $\markf$ is $\Q^{+}$-reachable from a $\Q^{+}$-marking $\marki$, then there exists a $\Q^{+}$-run $\rho$ which uses polynomially many data values and $\marki \xrightarrow{\rho}_{\Q^{+}} \markf$.
We first formalize the set of distinct data values associated with  $\X$-markings, data values \emph{used} in $\X$-runs and variables associated with a transition. 
\begin{definition}
\label{def:dval}
For $\mathcal{N} = \Net{P,T,F,\Var}$ a UDPN, $\X$-marking $\markm$, $t \in T$, and  $\X$-run $\rho = \run{c_i,t_i,\pi_i}{|\rho|}$, we define
\begin{enumerate}
    \item $\dval(\markm) = \{  \alpha \in \D \mid  ~\exists p \in P : \markm(p,\alpha) \neq 0 \}$. 
    \item $\dval(\rho) = \{  \alpha \in \D \mid ~ \exists i \leq |\rho|~ \exists x  \in \dval(t_i) : (\pi_i(x) = \alpha ) \}$.
    \item $\var(t) = \{ x\in \Var \mid ~ \exists p \in P ~: F(p,t)(x) \neq 0 \lor F(t,p)(x) \neq 0\}$.
\end{enumerate}
\end{definition}

With this we state the first main result of this paper, which  provides a bound on witnesses of $\Q,\Q^+$-reachability, and is proved in Section~\ref{sec:Bound}. 
\begin{theorem}
\label{thm:main2}
For $\X \in \{\Q,\Q^{+}\}$, if an $\X$-marking $\markf$ is $\X$-reachable from an initial $\X$-marking $\marki$, then there is an $\X$-run $\rho$ such that $\marki\xrightarrow{\rho}_{\X}\markf$
and $|\dval(\rho)| \leq |\dval(\marki)\cup \dval(\markf)|+1+\max_{t \in T}(|\var(t)|)$.
\end{theorem}
\indent Using the above bound, we
obtain a polynomial time algorithm for $\Q$-reachability, as detailed in Section~\ref{Sec:Q-reach}.
\begin{restatable}{theorem}{qreach}
\label{thm:qreach}
Given $\mathcal{N} = \Net{P,T,F,\Var}$ a UDPN and two $\Q$-markings \marki, \markf, deciding if $\markf$ is $\Q$-reachable from $\marki$ in $\mathcal{N}$ is in polynomial time. 
\end{restatable}

Finally, we consider continuous, i.e., $\Q^+$-reachability for UDPN. We adapt the techniques used for $\Q^+$-reachability of Petri nets without data from~\cite{DBLP:journals/fuin/FracaH15,DBLP:conf/lics/BlondinH17} to the setting with data, and obtain a characterization of $\Q^+$-reachability for UDPN  in Section~\ref{Sec:CharacterizingQ+-reachability}. Finally, in Section~\ref{Sec:Q+algorithm}, we show how the characterization can be combined with the above bound and compression techniques from \cite{DBLP:conf/lics/HofmanLT17} to obtain a polynomial sized system of linear equations with implications over $\Q^+$.  To do so, we require a slight transformation of the net which is described in Section~\ref{section:trans7.2}.

This  leads to our headline result, stated below.

\begin{theorem}[Continuous reachability for UDPN]\label{thm:main}
Given $\mathcal{N} = \Net{P,T,F,\Var}$ a UDPN and two $\Q^{+}$-markings \marki, \markf, deciding if $\markf$ is $\Q^{+}$-reachable from $\marki$ in $\mathcal{N}$ is in polynomial time. 
\end{theorem}

\noindent The rest of this paper is dedicated to proving these theorems. First, we present an equivalent formulation via matrices, which simplifies the technical arguments.

\section{Equivalent formulation via Matrices}
\label{sec:Matrices}
From now on, we restrict $\X$ to a symbol denoting $\Q$ or $\Q^+.$ We formulate the definitions presented earlier in terms of matrices, since defining object such as $\X$-marking as functions 
is intuitive to define but difficult to operate upon. 

In the following, we abuse the notation and use the same names for objects as well as matrices representing them. We remark that this is safe as all arithmetic operations on objects correspond to matching operations on matrices.

An $\X$-marking $\markm$ is a $P \times \D$ matrix $M$, where $ \forall p \in P,\forall \alpha \in \D, M(p,\alpha) = \markm(p,\alpha)$. As a \emph{finite representation}, we keep only a $P\times\dval(\markm)$ matrix of non-zero columns. 
For a transition $t \in T$, we represent $F(t,\bullet), F(\bullet,t)$ as $P \times \Var$ matrices. 
Note that $(t,\bullet)$ is not the position in the matrix, but is part of the name of the matrix; its entry at $(i,j)\in P\times \Var$ is 
given by $F(t,\bullet)(i,j)$.
For a place $p \in \row(F(t,\bullet))$, the row $F(t,\bullet)(p,\bullet)$ is a vector
in $\N^{\Var}$, given by an equation $F(\bullet, t)(p,\bullet)(x)=F(p,t)(x)$ for $p\in P, t\in T, x\in \Var.$
Similarly, $\Delta(t)$ is a $P \times \Var$ matrix with $\Delta(t)(p,x)=F(t,\bullet)(p,x)-F(\bullet,t)(p,x)$ for $t\in T, p\in P, \text{ and }x\in\Var.$
Although, both $\Delta(t)$ and $F(\bullet,t)$ are defined as $P\times \Var$ matrices, only the columns for variables in $\var(t)$ may be non-zero, so often we will iterate only over $\var(t)$ instead of $\Var$.

Finally, we capture a mode $\pi : \Var \to \D$ as a $\Var \times \D$ permutation matrix $\Pmat$. Although $\Pmat$ may not be a square matrix, we abuse notation and call them permutation matrices. $\Pmat$ basically represents assignment of variables in $\Var$ to data values just like $\pi$ does. An entry of 1 represents that the corresponding variable is assigned corresponding data value in mode $\pi$. Thus, for each mode $\pi : \Var \to \D$ there is a permutation matrix $\Pmat_{\pi}$, such that for all $x \in \Var$, $\alpha \in \col(\Pmat_{\pi})$, $\Pmat_{\pi}(x,\alpha) =  1$ if $\pi(x) = \alpha$, and $\Pmat_{\pi}(x,\alpha)=0$ otherwise.
Formulating a mode as a permutation matrix has the advantage that $\Delta(t)\circ \pi^{-1}$ is captured by $\Delta(t) \cdot \Pmat_{\pi}$ where $\Pmat_{\pi}$ can be represented as a sub-matrix of actual $\Pmat_{\pi}$, whose set of row indices is limited to the set of column indices of the matrix $\Delta(t)$. 

\begin{example}In the UDPN $\cN_1$ from
Example~\ref{ex:1}, the initial marking $\marki$ can be represented by the matrix $\marki$ below and the function $\Delta(t)$ by the matrix $\Delta(t)$ 
\[
  {{\marki}}=\ 
  \begin{blockarray}{ccccc}
red & blue & green & black &  \\
\begin{block}{(cccc) c}
  1 & 0 & 1 & 0 & \ \  p_1 \\
  0 & 1 & 0 & 0 & \ \ p_2\\
  2 & 0 & 0 & 0 & \ \ p_3\\
  1 & 1 & 0 & 0 & \ \ p_4\\
\end{block}
\end{blockarray}\quad
  {\Delta(t)}=
  \begin{blockarray}{cccc}
x & y & z &  \\
    \begin{block}{(ccc)c}
  0 & -1 & 0 &  \ \ p_1 \\
  -1 & 0 & 0 &  \ \ p_2\\
  0 & 2 & 0 &   \ \ p_3\\
  1 & 0 & 1 &   \ \ p_4\\
\end{block}
\end{blockarray}
\]
If we fire transition $t$ with the assignment $x=blue, y=green, z=black$, we get the following net depicted below (left), with marking $\markf$ (below center).
The permutation matrix corresponding to the mode of fired transition is given by $\Pmat$ matrix on the right. Note that the matrix $\markf - \marki$ is indeed the matrix $\Delta(t)\cdot \Pmat$.
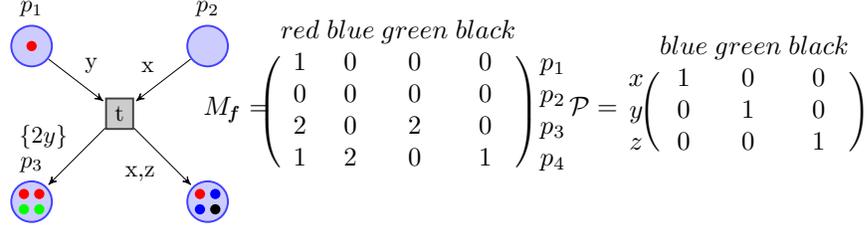
\begin{equation*}
\vcenter{
\scalebox{0.9}{
    \begin{tikzpicture}[node distance=1.3cm,>=stealth',bend angle=45,auto]
  \tikzstyle{place}=[circle,thick,draw=blue!75,fill=blue!20,minimum size=6mm]
  \tikzstyle{red place}=[place,draw=red!75,fill=red!20]
  \tikzstyle{transition}=[rectangle,thick,draw=black!75,
  			  fill=black!20,minimum size=4mm]

  \begin{scope}
    \node [place,label=$p_2$] (w1)[xshift=0mm,yshift=0mm]{};
     \node [transition] (t1) [left of=w1,,xshift=0mm,yshift=-10mm] {t}
      edge [pre] node{x} (w1);
    \node [place,colored tokens={red},label=$p_1$] (w2)[left of=t1,xshift=0mm,yshift=10mm]{}
    edge [post] node{y}(t1);
    \node [place,colored tokens={red,blue,blue,black}] (p3)[below of=w1,xshift=0mm,yshift=-10mm]{}
    
    edge [pre] node{x,z}(t1);
   
      \node [place,colored tokens={red,red,green,green},label=$p_3$] (p4)[below of=w2,xshift=0mm,yshift=-10mm]{}
      edge [pre] node{$\{2y\}$}(t1);
  \end{scope}
\end{tikzpicture}}}
%
{\hspace*{-9cm}
M_{\markf}=
  \begin{blockarray}{ccccc}
red & blue & green & black &  \\
\begin{block}{(cccc)c}
  1 & 0 & 0 & 0 & \ \ p_1 \\
  0 & 0 & 0 & 0 & \ \ p_2\\
  2 & 0 & 2 & 0 &  \ \ p_3\\
  1 & 2 & 0 & 1 &  \ \ p_4\\
\end{block}
\end{blockarray}
  \Pmat =
 \begin{blockarray}{cccc}
 & blue & green & black \\
    \begin{block}{c(ccc)}
 x \ \ &   1 & 0 & 0  \\
y \ \ &   0 & 1 &  0 \\
z \ \ &    0 & 0 &  1 \\
\end{block}
\end{blockarray}
}\\
\end{equation*}
\end{example}

Using the representations developed so far we can represent an $\X$-run $\rho$ as $\run{c_i,t_i,\Pmat_i}{|\rho|}$ 
 where $(c_i, t_i, \Pmat_i)$ denotes the $i^{th}$ $\X$-step fired with coefficient $c_i$ using transition $t_i$ with a mode corresponding to the permutation matrix $\Pmat_i$. 
 The sum of the matrices ($\sum_{i=1}^{|\rho|}c_i \Delta(t_i)\cdot \Pmat_i$) gives us the effect of the run i.e.
 $\Delta(\rho) = 
 \markf - \marki$ where 
 $\marki \xrightarrow{\rho}_{\X} \markf$. Effect of an $\X$-run $\rho$ on a data value $\alpha$ is $\Delta(\rho)(\bullet,\alpha) $. Also, for an $\X$-run $\rho =  \run{c_i,t_i,\Pmat_i}{|\rho|}$
, define $k{\rho} = \run{k{c_i},t_i,\Pmat_i}{|\rho|}$ where $k \in \X^{+}$.


\section{Bounding number of data values used in ${\Q,\Q^{+}}$-run} \label{sec:Bound}
 We now prove the first main result of the paper, namely, Theorem~\ref{thm:main2}, which shows a linear upper bound on the number of data values required in a $\Q^+$-run and a $\Q$-run. 
Theorem~\ref{thm:main2} is an immediate consequence of the following lemma, which states that if more than a linearly bounded number of data values are used in a $\Q$ or $\Q^+$ run, then there is another such run in which we use at least one less data value. 
\begin{lemma}\label{lemma:Q-dec}
Let $\X\in\{\Q,\Q^+\}$. If there exists an $\X$-run $\sigma$ such that $\marki\xrightarrow{\sigma}_{\X}\markf$ and $|\dval(\sigma)| > |\dval(\marki) \cup \dval(\markf)|+1+\max_{t \in T}(|\var(t)|)$, then there exists an $\X$-run $\rho$ such that $\marki \xrightarrow{\rho}_{\X} \markf$ and 
$|\dval(\rho)|\leq |\dval(\sigma)|-1$.
\end{lemma}
 By repeatedly applying this lemma, Theorem~\ref{thm:main2}  follows immediately. The rest of this section is devoted to proving this lemma. The central idea is to take any $\Q$ or $\Q^+$-run between $\marki$, $\markf$ and transform it to use at least one data value less. 

\subsection{Transformation of an $\X$-run}
The transformation which we call \emph{decrease}  is defined as a combination of two separate operations on an $\X$-run; we name them $\uniformize$ and $\replace$ and denote them by $\uniformyzing$ and $\reducing$ respectively. 
\begin{itemize}
    \item $\uniformize$ takes an $\X$-step and a non-empty set of data values $\setE$ as input and produces an $\X$-run, such that in the resultant run, the effect of the run for each data value in $\setE$ is equal.
    \item $\replace$ takes an $\X$-step, a single data value $\da,$ and a non-empty set of data values $\setE$ as input and outputs an $\X$-step which doesn't use data value $\da$.
\end{itemize}
The intuition behind the decrease operation is that we would like to take two data values $\da$ and $\beta$ used in the run such that effect on both of them is $\zero$ (they exists as the effect on every data value not present in the initial of final configuration is $\zero$) and replace usage of $\da$ by $\beta$. 
However, such a replacement can only be done if both data are not used together in a single step (indeed, a mode $\pi$ cannot assign the same data values to two variables). 
Unfortunately we cannot guarantee the existence of such  a $\beta$ that may replace $\da$ globally. We circumvent this by applying the $\replace$ operation separately for every step, replacing $\da$ with different data values in different steps.

But such a transformation would not preserve the effect of the run. To repair this aspect we uniformize i.e. guarantee that the final effect after replacing $\da$ by other data values is equal for every datum that is used to replace $\da$. As the effect on $\da$ was $\zero$ then if we split it uniformly it adds $\zero$ to effects of data replacing $\da$, which is exactly what we want. We now formalize this intuition below.
\subsubsection{The uniformize operator.}
By $\concatenate$ we denote an operator of concatenation of two sequences.
Although the data set $\D$ is unordered, the following definitions require access to an arbitrary but fixed  linear order on its elements. The definition of the $\uniformize$ operator needs another operator to act on an $\X$-step, which we call $\rotate$ and denote by $\rotating$. 

\begin{definition} \label{def:g-operation}
For a non-empty set of data values $\setE \subset \D$ and an $\X$-step, $\omega = (c,t,\Pmat)$, define $\rotating(\setE,\omega)= (c,t,\Pmat')$ where $\Pmat'$ is obtained from $\Pmat$ as follows.
\begin{itemize}
    \item $\forall \da \in \col(\Pmat) \setminus \setE$, $\Pmat'(\bullet,\da) = \Pmat(\bullet,\da)$.
    \item $\forall \da \in \setE $, $\Pmat'(\bullet,\da) = \Pmat(\bullet,next_{\setE}(\da))$, where $next_{\setE}(\da) =  \min(\{ \db \in \setE \mid  \db > \da  \})$ if\ \  $|\{ \db \in \setE \mid \db > \da  \}| > 0$ and $\min(\setE)$ otherwise. 
\end{itemize}
\end{definition}
For a fixed set $\setE$, we can repeatedly apply $\rotating(\setE,\bullet)$ operation on an $\X$-step, which we denote by $\rotating^k(\setE,\omega)$, where $k$ is the number of times we applied the operation (for example: $\rotating^2(\setE,\omega)=\rotating(\setE,(\rotating(\setE,\omega))$). 
\begin{definition} \label{def:uniformize}
For a non-empty set of data values $\setE \subset \D$ and an $\X$-step $\omega = (c,t,\Pmat)$, we define $\uniformize$ as follows 
\begin{center}
    $\uniformyzing(\setE,\omega) = \rotating^{0}(\setE,\frac{\omega}{|\setE|})~ \concatenate~ \rotating^{1}(\setE,\frac{\omega}{|\setE|}) ~ \concatenate ~\rotating^{2}(\setE,\frac{\omega}{|\setE|})~\concatenate~ ... ~\concatenate~ \rotating^{|\setE|-1}(\setE,\frac{\omega}{|\setE|}) $.
\end{center}
\end{definition}

An important property of uniformize is its effect on data values.
\begin{restatable}{lemma}{help}
\label{lem:help}
For a non-empty set of data values ${\setE}\subset \D$ and an $\X$-step $\omega = (c,t,\Pmat)$, $\marki \xrightarrow{\omega}_{\Q^+} \markf$,  if $\marki'\xrightarrow{\uniformyzing({\setE},\omega)} \markf'$, then 
\begin{enumerate}
    \item $\forall \da \in \dval( \omega)\backslash {\setE} $, $ \markf'(\bullet,\da)-\marki'(\bullet,\da)=\markf(\bullet,\da)-\marki(\bullet,\da)$
    \item $\forall \da \in {\setE},~, \markf'(\bullet,\da)-\marki'(\bullet,\da)=\frac{\sum_{\db \in {\setE}}(\markf(\bullet,\db)-\marki(\bullet,\db))}{|\setE|}$. 
\end{enumerate}
\end{restatable}
This lemma tells us the effect of the run on the initial marking is equalized for data values in ${\setE}$ by the $\uniformyzing$ operation, and is unchanged for the other data values. 

\subsubsection{The replace operator.}
To define the $\replace$ operator it is useful to introduce $swap_{\da,\db}(\Pmat)$ which exchanges  columns $\da$ and $\db$ in the matrix $\Pmat$.
\begin{definition}
For a set of data values $\setE$, an $\X$-step $\omega = (c,t,\Pmat)$, and $\da\not\in \setE$  we define $\replace$ as follows 
    \[ \reducing(\da,\setE,\omega) = \begin{cases} 
      (c,t,\Pmat) & \text{if } (\Delta(t)\cdot \Pmat)(\bullet,\da)= \zero   \\
      (c,t,swap_{\da,\db}(\Pmat)) &  \text{otherwise, } \text{where }\db \text{ is a smallest}\\ &
      \text{datum $\in\setE$ such that }{(\Delta(t)\cdot\Pmat)(\bullet,\db)=\zero}  
    \end{cases}
\]
\end{definition}
 After applying the $\replace$ operation $\alpha$ is no longer used in the run, which reduces the number of data values used in the run.
   Observe that $\replace$ can not be always applied to an $\X$-step. It requires a zero column labelled with an element from $\setE$ in the permutation matrix corresponding to the $\X$-step. 
    
    \subsubsection{The decrease transformation.} 
Now we are ready to define the final transformation on an $\X$-run between two markings which we call $\decrease$ and denote by $\decreasing$. 

\begin{definition}\label{def:decrease}
For two $\X$-markings $\marki$, $\markf$, and an $\X$-run $\sigma$ such that $\marki \xrightarrow{\sigma}_{\X} \markf$ and $|\dval(\sigma)| > |\dval(\marki) \cup \dval(\markf)|+1+\max_{t \in T}(|\var(t)|)$, let $\{\da\} \cup\setE=\dval(\sigma) \setminus (\dval(\marki) \cup \dval(\markf))$ and $\da\not\in \setE$. 
We define $\decrease$ by, $\decreasing(\setE,\da,\sigma) =$
\begin{equation*}
 \uniformyzing(\setE,\reducing(\da,\setE,\sigma(1)))~\concatenate~\uniformyzing(\setE,\reducing(\da,\setE,\sigma(2)))~
\concatenate~...~\concatenate~\uniformyzing(\setE,\reducing(\da,\setE,\sigma(|\sigma|))).      
\end{equation*}
where $\sigma(j)$ denotes the $j^{th}$ $\X$-step of $\sigma$.
 \end{definition}

Observe that the required size of $\dval(\sigma)$ guarantees existence of a $\beta\in \setE$ which can be replaced with $\da$, for every application of the $\reducing$ operation. Note that the exchanged data value $\beta$ could be different for each step.
Finally, we can analyze the $\decrease$ transformation and show that if the original run allows for the $\decrease$ transformation (as given in the above definition), then after the application of it, the resulting sequence of transitions is a valid run of the system.

\begin{lemma}\label{lem:decr}
 Let $\sigma$ be an $\X$-run such that $\marki \xrightarrow{\sigma}_{\X} \markf$ and $|\dval(\sigma)| > |\dval(\marki) \cup \dval(\markf)|+1+\max_{t \in T}(|\dval(t)|)$. Let $\da \in \dval(\sigma) \setminus (\dval(\marki) \cup \dval(\markf))$ and $\setE = \dval(\sigma) \setminus (\dval(\marki) \cup \dval(\markf) \cup \{\da\})$. Then for $\rho=\decreasing(\setE, \da, \sigma)$, we obtain 
 $\marki\xrightarrow{\rho}_{\X} \markf.$
\end{lemma}
\begin{proof}
Suppose $\sigma=\sigma_1\sigma_2\ldots\sigma_l$ where each
$\sigma_j=(c_j, t_j, \Pmat_j)$, for $1\leq j\leq l$ is an $\X$-step. Then $\rho=\rho_1\concat \ldots\concat \rho_l$, where each $\rho_j$ is an $\X$-run defined by $\rho_j=\uniformyzing({\setE},\reducing(\da,\setE,\sigma_j))$. It will be useful to identify intermediate $\X$-markings \begin{gather}\marki=\markm_0\xrightarrow{\sigma_1}_{\X}\markm_1 \xrightarrow{\sigma_2}_{\X}\markm_2
\xrightarrow{\sigma_3}_{\X}\ldots\xrightarrow{\sigma_l}_{\X}\markm_l=\markf\\
\marki=\markm_o'\xrightarrow{\uniformyzing({\setE},\reducing(\da,\setE,\sigma_1))}_{\Q}\markm_1' \xrightarrow{\uniformyzing({\setE},\reducing(\da,\setE,\sigma_2))}_{\Q}\markm_2'
\ldots
\xrightarrow{\uniformyzing({\setE},\reducing(\da,\setE,\sigma_l))}_{\Q}\markm'_l=\markf'\end{gather}

\noindent We split the proof: first we show that $\markf=\markf'$ and then $\rho$ is $\X$-fireable from $\marki.$

{\bf Step 1: Showing that the final markings reached are the same.}
We prove a stronger statement which implies that $\markf=\markf'$, namely:
\begin{restatable}{myclaim}{claimClaimOne}\label{claim:bnd} For all $0\leq j\leq l$ 
\begin{enumerate}
    \item $\markm_j'(\bullet,\da)=\zero$
    \item $\forall \gamma \in $ $\dval(\marki)\cup \dval(\markf)$,  $\markm_j'(\bullet,\gamma)=\markm_j(\bullet,\gamma)$
    \item $\forall \gamma \in $ ${\setE}$ $ \markm_j'(\bullet,\gamma)=\frac{1}{|\setE|}\left(\sum_{\delta\in{\setE}\cup \{\da\}}  \markm_j(\bullet,\delta)\right).$
    \end{enumerate}
\end{restatable}
The proof is obtained by induction on $j$, and is a resulting computations as detailed in Appendix~\ref{app:sec5}. Intuitively, point 1 holds as we shift effects on $\da$ to $\beta$-s, point 2 holds as the transformation does not touch $\gamma\in \dval(\marki)\cup\dval(\markf).$ The last most complicated point follows from the fact that the number of tokens consumed and produced along each $\xrightarrow{\uniformyzing({\setE},\reducing(\da,\setE,\sigma_j))}$ is the same as for $\sigma_j$, but uniformized over $\setE$.

{\bf Step 2: Showing that $\rho$ is an $\X$-run.}
If $\X=\Q$ then the run $\rho$ is fireable, as any $\Q$-run is fireable, so in this case this step is trivial. The case when $\X=\Q^+$ is more involved. As we know from claim \ref{claim:bnd} , each $m_j'$ is a $\Q^+$-marking, so it suffices to prove that for every $j$, $\markm_j'\xrightarrow{\uniformyzing({\setE},\reducing(\da,\setE,\sigma_j))}_{\Q^{+}}\markm_{j+1}'$.
Consider a data vector of tokens consumed along the $\Q^+$-run $\uniformyzing({\setE},\reducing(\da,\setE,\sigma_j))$. If we show that it is smaller than or equal to $\markm_j'$ (component-wise), then we can conclude that $\uniformyzing({\setE},\reducing(\da,\setE,\sigma_j))$ is indeed $\Q^+$-fireable from $\markm_j'$.  To show this, we examine the consumed tokens for each datum $\gamma$ separately. There are three cases:
\begin{itemize}
    \item[(i)] $\gamma=\da$. For this case, every step in $\uniformyzing({\setE},\reducing(\da,\setE, \sigma_j))$ does not make any change on $\da$ so tokens with data value $\da$ are not consumed along the $\Q^+$-run $\uniformyzing({\setE},\reducing(\da,\setE,\sigma_j))$.
    \item [(ii)] $\gamma\in\dval(\marki)\cup\dval(\markf)$. This is similar to the above case.  Consider any data value $\gamma \in (\dval(\sigma)\backslash {\setE})\setminus\{\da\}$.    Since $\gamma$ does not change on $\rotate$ operation, the $\uniformyzing$ operation causes each $\Q$-step in $\uniformyzing({\setE},\reducing(\da,\setE, \sigma_j))$ to consume $\frac{1}{|\setE|}$ of the tokens with data value $\gamma$ consumed when $\sigma_j$ is fired. 
 This is repeated $|\setE|$ times and hence the vector of tokens with data value $\gamma$ consumed along $\uniformyzing({\setE},\reducing(\da,\setE, \sigma_j))$ is equal  to the vector of tokens with value $\gamma$ consumed by step  $\sigma_j$.  But we know that, it is smaller than $\markm_j(\bullet,\gamma)$ and concluding smaller than $\markm_j'(\bullet,\gamma)$. The last inequality is true as $\markm_j(\bullet,\gamma)=\markm_j'(\bullet,\gamma)$ according to Claim~\ref{claim:bnd}.
\item[(iii)] $\gamma\in \setE$.  Let $\omega$ be a triple $(c_j,F(\bullet,t_j), \Pmat_j)$ where $(c_j,t_j, \Pmat_j)=\sigma_j.$  $\omega$ simply describes tokens consumed by $\sigma_j.$
We slightly overload the notation and treat a triple $\omega$ like a step, where $F(\bullet,t_j)$ represents a transition "\_" for which 
$F(\bullet,\_)=F(\bullet,t_j)$ and $F(\_,\bullet)$ is a zero matrix.
We calculate the vector of consumed tokens with data value $\gamma$ as follows: $  consumed(\bullet,\gamma)=$
$$\frac{1}{|\setE|}\sum_{k=0}^{|\setE|-1} \Delta(\rotating^{k}(\setE,\reducing(\da,\setE,\omega)))(\bullet,\gamma)= 
  \frac{1}{|\setE|}\sum_{k=0}^{|\setE|} \Delta(\rotating^{k}({\setE\cup\{\da\}},\omega))(\bullet,\gamma)$$
 the first equality is from definition and the second by the $\replace$ operation,
 $$=\frac{c_j}{|\setE|}\sum_{k=0}^{|\setE|} (\rotating^k( F(\bullet,t_j)\cdot \Pmat_j))(\bullet,\delta)= 
 \frac{c_j}{|\setE|} \sum_{\delta\in\setE\cup\{\da\}} (F(\bullet, t_j)\cdot \Pmat_j)(\bullet,\delta).
 $$
 Further, observe that as $\sigma_j$ can fired in $\markm_j$ 
 $$c_j(F(\bullet,t_j)\cdot \Pmat_j)(\bullet,\delta)\leq   \markm_j(\bullet,\delta) \text{ for all }\delta\in \D,$$
summing up over $\delta\in\setE\cup\{\da\}$ and multiplying with $\frac{1}{|\setE|}$ we get 
$$\frac{1}{|\setE|}c_j\sum_{\delta\in\setE\cup\{\da\}} (F(\bullet,t_j)\cdot \Pmat_j)(\bullet,\delta) \leq   \frac{1}{|\setE|}\sum_{\delta\in\setE\cup\{\da\}} \markm_j(\bullet,\delta)= \markm_j'(\delta,\gamma),$$
where the last equality comes from Claim~\ref{claim:bnd} point 3. Combining inequalities we get
$consumed(\bullet,\gamma)\leq \markm_i'(\bullet,\gamma)$.
\end{itemize}
\end{proof}

\begin{proof}[of Lemma~\ref{lemma:Q-dec}]
Now the proof of Lemma~\ref{lemma:Q-dec} (and hence Theorem~\ref{thm:main2}) follow immediately, since we can use the $\decrease$ transformation, to decrease the number of data values required in an $\X$-run. We simply take
$\da\in\dval(\sigma)\setminus(\dval(\marki)\cup\dval(\markf))$ and $\setE=\dval(\sigma)\setminus(\dval(\marki)\cup\dval(\markf))\setminus\{\da\}.$
Next, let $\rho=\decreasing({\setE},\da,\sigma)).$ Due to Lemma~\ref{lem:decr} we know that $\marki\xrightarrow{\rho}_{\X}\markf$.
Moreover, observe that 
$\dval(\rho)\subseteq\dval(\sigma)$. But in addition,
$\da\not\in \dval(\rho)$ as due to the one of properties of the $\decrease$ operation $\da$ does not participate in the run $\rho$.
So $\dval(\rho)\subset\dval(\sigma).$ Therefore $|\dval(\rho)|\leq |\dval(\sigma)|-1$.
\end{proof}

\section{$\Q$-reachability is in \PTIME} \label{Sec:Q-reach}

\indent We recall the definition of histograms from \cite{DBLP:conf/lics/HofmanLT17}.
\begin{definition} \label{def:histogram} 
A histogram $M$ of order $q \in \mathbb{Q}$ is a $\Var \times \D$ matrix having non-negative rational entries 
such that, 
\begin{enumerate}
    \item $\sum_{\alpha \in \col(M)}M(x,\alpha) = q$ for all $x \in \row(M)$.
    \item $\sum_{x \in \row(M)}M(x,\alpha) \leq q$ for all $\alpha \in \col(M)$.
\end{enumerate}
\end{definition}
A permutation matrix is a histogram of order 1. 
 
{We now state two properties of histograms in the following lemma. We say that a histogram of order $a$ is an \emph{[$a$]-histogram} if the histogram has only $\{0,a\}$ entries.}  

{
\begin{restatable}{lemma}{histadd}\label{lemma:hist add}
Let $H,H_1,H_2,..,H_n$ be histograms of order $q,q_1,q_2,...,q_n$ respectively and of same row dimensions then
    (i) $\sum_{i=1}^{n}H_i$ is a histogram of order $\sum_{i}^{n}q_i$,
    (ii) $H$ can be decomposed as a sum of [$a_i$]-histograms such that $\sum_{i}a_i = q$.
\end{restatable}
}

Using histograms we define a representation $\Hist(\rho)$ for an $\X$-run $\rho$, which captures $\Delta(\rho)$. 
From an $\X$-run $\rho = \run{c_j,t_j,\Pmat_j}{|\rho|}$ we obtain $\Hist(\rho)$ as follows. For all transitions $t \in T $, define the set $I_t = \{ j \in [1..|\rho|]|~  t_j = t \}$. Then calculate the matrix
 $ H_t = \sum_{i \in I_t} c_i \Pmat_i$. Observe that since permutation matrices are histograms and  histograms are closed under scalar multiplication and addition, $H_t$ is a histogram. If $I_t$ is empty, then $H_t$ is simply the null matrix.
We define $\Hist(\rho)$ as a mapping from $T$ to histograms such that $t$ is mapped to $H_t$.

Analogous to an $\X$-run we can represent $\Hist(\rho)$ simply as $\{(t_j,H_{t_j})\}$, unlike an $\X$-run we don't indicate the length of the sequence since
it is dependent on the net and not the individual run itself. 
 \begin{proposition}
 Let $\mathcal{N} = \Net{P,T,F,\Var}$ be a UDPN, $\marki,\markf$ \ $\X$-markings, and $\sigma$ an $\X$-run such that 
 $\marki \xrightarrow{\sigma}_{\X} \markf$. Then for each $t\in T$ there exists $H_t$ such that:
 \begin{enumerate}\label{prop:histograms}
     \item $\markf-\marki=\sum_{t\in T} \Delta(t)\cdot H_t,$
     \item $\col(H_t)\subseteq \dval(\sigma)$ for every $t\in T.$
 \end{enumerate}
 \end{proposition}

\noindent {\textbf{A PTime Procedure.}}
We start by observing that from 
any $\Q$-marking $\marki$, every $\Q$-step $(c,t,\Pmat)$ is fireable and every $\Q$ run is fireable.
This follows from the fact that rationals are closed under addition, thus $\marki + c\cdot F(\bullet,t) \cdot \Pmat$ is a marking in $\mathcal{M}_{\Q}$.  
Thus if we have to find a $\Q$-run $\rho = \run{c_j,t_j,\Pmat_j}{|\rho|}$ between two $\Q$-markings, $\marki,\markf$ it is sufficient to ensure that $\markf -\marki = \sum_{j=1}^{|\rho|}c_j\Delta(t_j)\cdot\Pmat_j$. Thus for a $\Q$-run all that matters is the difference in markings caused by the $\Q$-run which is captured succinctly by $\Hist(\rho) = \{t_j,H_{t_j}\}$. This brings us to our characterization of $\Q$-run.

\begin{restatable}{lemma}{QCharacterization}
\label{lemma:Q-characterization}
Let $\mathcal{N} = \Net{P,T,F,\Var}$ be a UDPN, a marking $\markf$ is $\Q$-reachable from $\marki$ iff there exists set $\setE$ of size bounded by 
$|\setE|\leq |\dval(\marki)\cup\dval(\markf)|+1+\max_{t \in T}(|\var(t)|)$ and a histogram $H_t$ for each $t \in T$ such that $\markf - \marki = \sum_{t\in T} \Delta(t)\cdot H_t $ and $\forall t \in T ~\col(H_t)\subseteq \setE.$ 
\end{restatable}
Using this characterization we can write a system of linear inequalities to encode the condition of Lemma \ref{lemma:Q-characterization}. 
Thus, we obtain our second main result, with detailed proofs in the Appendix~\ref{app:sec6}.
\qreach*

\section{$\Q^+$-reachability is in \PTIME}
Finally, we turn to $\Q^+$-reachability for UDPNs and  
to the proof of Theorem~\ref{thm:main}. At a high level, the proof is in three steps. We start with a characterization of $\Q^+$-reachability in UDPNs.

Then we present a polytime reduction of the continuous reachability problem to the same problem but for a special subclass of UDPN, called loop-less nets.
Finally, we present how to encode the characterization for loop-less nets into a system of \emph{linear equations with implications} to obtain a polytime algorithm for continuous reachability in UDPNs.

\removed{However, we cannot encode this characterization directly into linear equations. Instead in the subsequent subsection, we reduce the problem to a special subclass of UDPN, called loop-less nets, with an equivalent continuous reachability problem. 
Finally, in the final and most complicated step, we will see how to encode the characterization for loop-less nets into a system of \emph{linear equations with implications} and exploit the bound proved in Theorem~\ref{thm:main2} to obtain a polytime algorithm for continuous reachability in UDPNs.}

\subsection{Characterizing $\Q^+$-reachability}\label{Sec:CharacterizingQ+-reachability}

We begin with a definition. For an $\X$-run we introduce the notion of the pre and post sets of $\X-$run. 
For an $\X$-run, $\rho = \run{c_i, t_i,\Pmat_i}{|\rho|}$
we define $Pre(\rho) = \{(\da,p) |~ \exists~ t_i, \exists~ x : F(p,t_i)(x) < 0 \wedge \Pmat_i(x,\da) = 1\}$. 
We also define $Post(\rho) = \{(\da,p) |~ \exists~ t_i, \exists~ x : F(t_i,p)(x) > 0 \wedge \Pmat_i(x,\da) = 1\}$. 
Intuitively, $Pre(\rho)/Post(\rho)$ denote the set of $(\da,p)$ (data value,place) pairs 
describing tokens that are consumed/produced by the run $\rho$.

Throughout this section, by a marking we denote a $\Q^+$-marking.
\begin{restatable}{lemma}{lemmaFinalCharac}
\label{lemma:final charac}
Let $\mathcal{N}=\Net{P,T,F, \Var}$ be an UDPN and $\marki,\markf$ are markings. For any $\Q^{+}$-run $\sigma$ such that $\marki \xrightarrow{\sigma}_{\Q^{+}} \markf$ there exist markings $\marki'$ and $\markf'$ (possibly on a different run) such that 
\begin{enumerate}
\item $\marki'$ is $\Q^+$-reachable from $\marki$ in at most $|P|\cdot |\dval(\sigma)|$ $\ \Q^+$-steps
\item There is a run $\sigma'$ such that $\dval(\sigma')\subseteq \dval(\sigma)$ and $\marki'\xrightarrow{\sigma'}_{\Q}\markf'$
\item $\markf$ is $\Q^+$-reachable from $\markf'$ in at most $|P|\cdot |\dval(\sigma)|$         \ \  $\Q^+$-steps
    \item $\forall (p,\da) \in Pre(\sigma'), \marki'(p,\da) > 0$ 
    \item $\forall (p,\da) \in Post(\sigma'), \markf'(p,\da) > 0$ 
\end{enumerate}
\end{restatable}
\begin{restatable}{remark}{remFinalCharac}\label{rem:FinalCharac}
If in conditions 1 and 3 we drop the requirement on the number of steps then the five conditions still imply continuous reachability.
\end{restatable}
Note that if there exist markings $\marki'$ and $\markf'$ and $\Q^+$ runs $\rho$, $\rho'$, $\rho''$ such that $\marki \xrightarrow{\rho}_{\Q^{+}} \marki', \marki' \xrightarrow{\rho'}_{\Q^{+}} \markf', \markf' \xrightarrow{\rho''}_{\Q^{+}} \markf$  then there is a $\Q^+$ run $\sigma$ such that $\marki \xrightarrow{\sigma}_{\Q^{+}} \markf$.The above characterization and its proof are obtained by adapting to the data setting, the techniques developed for continuous reachability in Petri nets (without data) in~\cite{DBLP:journals/fuin/FracaH15} and ~\cite{DBLP:conf/lics/BlondinH17}. Details are in Appendix~\ref{app:sec71}. 


\subsection{Transforming UDPN to \loopless\ UDPN}
\label{section:trans7.2}

For a UDPN $\mathcal{N}=\Net{P,T,F,\Var}$, we construct a UDPN $N'$ which is polynomial in the size of $N$ and for which the $\Q^+$-reachabilty problem is equivalent.

We define $PrePlace(t)=\{p \in P |  \exists v \in \Var\ s.t.\ F(p,t)(v)>0 \}$ and $PostPlace(t)=\{p \in P|  \exists v \in \Var\ s.t.\ F(t,p)(v)>0 \}$, where $t\in T$.
The essential property of the transformed UDPN is that for every transition the sets of PrePlace and PostPlace do not intersect. 
%
A UDPN $N=\Net{P,T,F, \Var}$ is said to be \emph{\loopless}\ if for all $t\in T$, $PrePlace(t) \cap PostPlace(t)=\emptyset.$

Any UDPN can easily be transformed in polynomial time into a \loopless\ UDPN such that $\Q^+$-reachability is preserved, 
by doubling the number of places and adding intermediate transitions. Formally, For every net $N$ and two markings 
$\marki,\markf$
in polynomial time one can construct a \loopless\ net $N'$ and two markings $\marki',\ \markf'$ such that
$\marki\xrightarrow{}_{\Q^+}\markf$ in the net $N$ iff $\marki'\xrightarrow{}_{\Q^+}\markf'$ in $N'.$
The proof of this statement is formalized in Section~\ref{section:transformation} in the Appendix, along with examples and transformation.
Now, the following lemma which describes a property of \loopless\ nets will be crucial for our reachability algorithm:

\begin{restatable}{lemma}{Qstep}
\label{Q-step}
In a \loopless\ net, for markings $\marki$,  $\markf$, if there exist a histogram $H$, and a transition t $\in$ T such that $\marki+\Delta(t)\cdot H=\markf$, then there exist a $\Q^+$-run $\rho$  such that $\marki \xrightarrow{\rho}_{\Q^+}\markf$.
\end{restatable}

\subsection{Encoding  $\Q^+$-reachability as linear equations with implications}\label{Sec:Q+algorithm}
Linear equations with implications are defined exactly as we use it in \cite{DBLP:conf/concur/HofmanL18} but they were introduced in \cite{DBLP:conf/lics/BlondinH17}.
We also call a system of linear equations with implications a $\implies$ system.
A
$\implies$
-system
is a finite set
of linear inequalities, all over the same variables, plus a finite set of
implications of the form
$x>0\implies y>0$,
where
$x,y$
are variables appearing in the linear inequalities.
\begin{lemma}\label{lemma:sol}
\cite{DBLP:conf/lics/BlondinH17} The $\Q^+$ solvability problem for a $\implies$ system is in $\PTIME$.
\end{lemma}
Our aim here will be to reduce the $\Q^+$-reachability problem to checking the solvability of a system of linear equations with implications, using the characterization of the problem established in 
Lemma~\ref{lemma:final charac}.
\begin{lemma}\label{final}
$\Q^+$-reachability in a  UDPN $N=\Net{P,T,F, \Var}$ between markings $\marki,\markf$ can be encoded as a set of linear equations with implications in P-time.
\end{lemma}
\begin{proof}
As mentioned in Subsection~\ref{section:trans7.2} , without loss of generality we may assume that UDPN $N$ is \loopless.
 Invoking Theorem~\ref{thm:main2} w.l.o.g we can assume that the $\Q^{+}$-run $\sigma$ uses at most $|\dval(\marki)\cup \dval(\marki)| + 1+\max_{t \in T}(|\var(t)|) $ data values, call $\setY$ the set of data vales used by $\sigma$.

As we need to describe several linear constraints we present them in terms of matrix multiplication. We use a word "array" instead of a matrix whenever
we mean a table with variables instead of constants. To encode the conditions of lemma \ref{lemma:final charac} as equations, we introduce markings $\marki'$ and $\markf'$ (they are used to represent the intermediate markings in Lemma \ref{lemma:final charac}). $\marki'$ and $\markf'$ are arrays of variables indexed with $P\times \setY$. As they should be evaluated to $\Q^+$-markings we introduce inequalities $\marki'\geq 0\text{ and }\markf'\geq 0.$ Then it is left to encode the conditions of Lemma~\ref{lemma:final charac} as linear equations, which we do in two steps.

\noindent \textbf{Encoding of Conditions 2, 4, 5 of lemma \ref{lemma:final charac}}
 \begin{itemize}
     \item We first encode Cond. 2 of lemma \ref{lemma:final charac} (the linear equation representing $\Q$-reachability w.r.t $\marki',\markf'$) as 
     $\markf'-\marki'=\sum_{i=1}^{|T|} \Delta(t_i)\cdot h_i\text,$ 
     where $h_i$ are arrays of variables of dimension $ \Var \times \setY$ to represent histograms, since all non-zero columns appear in $\setY$.
     To guarantee that each array $h_i$ is encoding a histogram  we add equations encoding Conditions 1 and 2 from Definition~\ref{def:histogram}. We use the following notation: variable $h_i[r][\da]$ is used to indicate the entry in row $r$ and column $\da \in \setY$ of the histogram array $h_i$. 
     Now, each entry is $h_i[r][\da]\geq 0$
     and for every $\da,r, i$, $$\sum_{\da\in \setY} h_i[r][\da] =\sum_{\da\in \setY} h_i[1][\da]\text{ and }\sum_{r\in\Var} h_i[r][\da]\leq \sum_{\db\in\setY} h_i[1][\db].$$
     \item To encode the conditions $4,5$ of lemma \ref{lemma:final charac} , we will need to allow for \\ implication relation between variables.   
     Therefore, we add the constraints
     \linebreak $\forall t_i\in T,~\forall p\in P ,~\forall r\in \Var, ~\forall \da\in \setY,$  
     \begin{equation*}
        \begin{split}
     (4)& F(p,t_i)(r)<0 ~ \land ~  h_i[r][\da] >0 ~ \implies ~ \marki'(p,\da) > 0 \\
     (5)& F(t_i,p)(r) > 0 ~\land~  h_i[r][\da] >0 ~\implies~ \markf'(p,\da) > 0 
     \end{split}
     \end{equation*}
     This set of implications ensures that if for a $\Q$-run $\sigma'$, $(p,d)\in Pre(\sigma')$, then $\marki'(p,d)>0$ and similarly for the post-set.
 
 \end{itemize}

\noindent \textbf{Encoding of conditions 1 and 3 of Lemma \ref{lemma:final charac}:}\\
 As these conditions are symmetric we explain in detail only the encoding of Condition 1.
 Knowing, $\dval(\sigma)\subseteq \setY$ we may bound the number of transitions from $\marki$ to $\marki'$ by $B=|P|\cdot|\setY|=|P|\cdot \left(|\dval(\marki) \cup \dval(\markf)|+1+\max_{t \in T}(|\var(t)|)\right).$
 The first problem in trying to encode a run here is that, we don't know the exact order on which transitions of $\sigma$ will be taken, the second is that we don't know the precise instantiation of them. We handle both problems by over-approximating reachability via at most $B$ steps by a reachability via runs in following schema:
 $$ \left\lbrack(t_1,h_{1,1})(t_2,h_{2,1})\ldots (t_{|T|},h_{|T|,1})\right\rbrack
 \ldots\left\lbrack (t_1,h_{1,B})
 \ldots (t_{|T|},h_{|T|,B})\right\rbrack$$
 where $h_{i,j}$ are histograms with columns from the set $\setY$ and the expression $(t_i,h_{i,j})$ denotes any $\Q^+$-run that uses only a single transition $t_i$. 
 To see that it is an over-approximation it suffices to see that any run of length at most $B$ can be performed within the schema. The mentioned over-approximation is sufficient for us due to Remark~\ref{rem:FinalCharac}. The $j^{th}$ step from the run can be found in the $j^{th}$ block $\left\lbrack(t_1,h_{1,j})\ldots (t_{|T|},h_{|T|,j})\right\rbrack$, histograms of all unnecessary transitions are instantiated to zero. 
 
 Having above we describe $\Q^+$-reachability within this schema restricted to data values from $\setY$. We do it by introducing sets of arrays describing configurations
 $\marki=\marki_0,\marki_1,\marki_2,\ldots,\marki_{B\cdot |T|}=\marki'$ between runs $(t_i,h_{i,j}).$ 
 Further, for all $0\leq i< B\cdot |T|$ we add equations  
 $\marki_i+\Delta(t_j)\cdot h_{j,k}=\marki_{i+1}$ where $i=(j-1)+(k-1)\cdot |T|$ and necessary equations
 guaranteeing $h_{j,k}$ to be histograms (as done for $h_i$ above). The described system is of polynomial size and correctly captures $\Q$-reachability.
 
 The last missing bit is to restrict solutions as we want to express only $\Q^+$-reachability.
  Of course all of $\marki_i$ should be non-negative so we add equations
 $\marki_i\geq 0\ \forall i\leq B\cdot |T|.$
 This suffices to capture $\Q^+$-reachability. Indeed, each of $\Q$-runs between $\marki_i$ and $\marki_{i+1}$ is using a single transition, and from Lemma~\ref{Q-step} we get that they are fireable $\Q^+$-runs.

 Thus, we have correctly described $\Q^+$ reachability via the schema from $\marki$ to $\marki'$.
   \end{proof}
Finally, we obtain Theorem \ref{thm:main} as a consequence of  Lemma \ref{lemma:sol} and Lemma~\ref{final}. 

\section{Conclusion}
\label{sec:conclusion}
In this paper, we provided a polynomial time algorithm for continuous reachability in UDPN, matching the complexity for Petri nets without data. This is in contrast to problems such as discrete coverability, termination, where Petri nets with and without data differ enormously in complexity, and to (discrete) reachability, whose decidability is still open for UDPN.   
As future work, we aim to implement the continuous reachability algorithm developed here, to build the first tool for discrete coverability in UDPN on the lines of what has been done for Petri nets without data. The main obstacle will be performance evaluation due to lack of benchmarks for UDPNs. Another interesting avenue for future work would be to tackle continuous reachability for Petri nets with ordered data, which would allow us to analyze continuous variants of Timed Petri nets and so on. 
\bibliographystyle{unsrt}
\bibliography{ref}

\begin{thebibliography}{10}

\bibitem{WSTS-everywhere}
Alain Finkel and Philippe Schnoebelen.
\newblock Well-structured transition systems everywhere!
\newblock {\em Theor. Comput. Sci.}, 256(1-2):63--92, 2001.

\bibitem{DBLP:journals/tcs/Rackoff78}
Charles Rackoff.
\newblock The covering and boundedness problems for vector addition systems.
\newblock {\em Theor. Comput. Sci.}, 6:223--231, 1978.

\bibitem{DBLP:conf/stoc/Kosaraju82}
S.~Rao Kosaraju.
\newblock Decidability of reachability in vector addition systems (preliminary
  version).
\newblock In {\em Proceedings of the 14th Annual {ACM} Symposium on Theory of
  Computing, May 5-7, 1982, San Francisco, California, {USA}}, pages 267--281,
  1982.

\bibitem{DBLP:conf/lics/LerouxS15}
J{\'{e}}r{\^{o}}me Leroux and Sylvain Schmitz.
\newblock Demystifying reachability in vector addition systems.
\newblock In {\em 30th Annual {ACM/IEEE} Symposium on Logic in Computer
  Science, {LICS} 2015, Kyoto, Japan, July 6-10, 2015}, pages 56--67, 2015.

\bibitem{Lipton}
E.~Cardoza, Richard~J. Lipton, and Albert~R. Meyer.
\newblock Exponential space complete problems for {P}etri nets and commutative
  semigroups: Preliminary report.
\newblock In {\em Proceedings of the 8th Annual {ACM} Symposium on Theory of
  Computing, May 3-5, 1976, Hershey, Pennsylvania, {USA}}, pages 50--54, 1976.

\bibitem{Czerwinski}
Wojciech Czerwinski, Slawomir Lasota, Ranko Lazic, J{\'{e}}r{\^{o}}me Leroux,
  and Filip Mazowiecki.
\newblock The reachability problem for {P}etri nets is not elementary (extended
  abstract).
\newblock {\em CoRR}, abs/1809.07115, 2018.

\bibitem{DBLP:journals/jcsc/Aalst98}
Wil M.~P. van~der Aalst.
\newblock The application of {P}etri nets to workflow management.
\newblock {\em Journal of Circuits, Systems, and Computers}, 8(1):21--66, 1998.

\bibitem{DBLP:conf/ac/Esparza96}
Javier Esparza.
\newblock Decidability and complexity of {P}etri net problems - an
  introduction.
\newblock In {\em Lectures on {P}etri Nets {I:} Basic Models, Advances in
  {P}etri Nets, the volumes are based on the Advanced Course on Petri Nets,
  held in Dagstuhl, September 1996}, pages 374--428, 1996.

\bibitem{Desel:1995:FCP:207572}
J\"{o}rg Desel and Javier Esparza.
\newblock {\em Free Choice {P}etri Nets}.
\newblock Cambridge University Press, New York, NY, USA, 1995.

\bibitem{Continuous-original}
David R. and Alla H.
\newblock Continuous {P}etri nets.
\newblock In {\em Proceddings of the 8th European Workshop on Application and
  Theory of Petri Nets, Zaragoza, Spain, 1987}, page 275–294, 1987.

\bibitem{DBLP:journals/fuin/FracaH15}
Est{\'{\i}}baliz Fraca and Serge Haddad.
\newblock Complexity analysis of continuous {P}etri nets.
\newblock {\em Fundam. Inform.}, 137(1):1--28, 2015.

\bibitem{DBLP:conf/lics/BlondinH17}
Michael Blondin and Christoph Haase.
\newblock Logics for continuous reachability in {P}etri nets and vector
  addition systems with states.
\newblock In {\em 32nd Annual {ACM/IEEE} Symposium on Logic in Computer
  Science, {LICS} 2017, Reykjavik, Iceland, June 20-23, 2017}, pages 1--12,
  2017.

\bibitem{DBLP:journals/automatica/DavidA94}
Ren{\'{e}} David and Hassane Alla.
\newblock Petri nets for modeling of dynamic systems: {A} survey.
\newblock {\em Automatica}, 30(2):175--202, 1994.

\bibitem{Alla1998ContinuousAH}
Hassane Alla and Ren{\'e} David.
\newblock Continuous and {Hybrid Petri Nets}.
\newblock {\em Journal of Circuits, Systems, and Computers}, 8:159--188, 1998.

\bibitem{ColoredPetriNets}
Kurt Jensen.
\newblock Coloured {P}etri nets - preface by the section editor.
\newblock {\em {STTT}}, 2(2):95--97, 1998.

\bibitem{Wang1998}
Jiacun Wang.
\newblock {\em Time {P}etri Nets}, pages 63--123.
\newblock Springer US, Boston, MA, 1998.

\bibitem{Abdulla:2001:TPN:647747.734218}
Parosh~Aziz Abdulla and Aletta Nyl{\'e}n.
\newblock Timed petri nets and bqos.
\newblock In {\em Proceedings of the 22Nd International Conference on
  Application and Theory of Petri Nets}, ICATPN '01, pages 53--70, London, UK,
  UK, 2001. Springer-Verlag.

\bibitem{Nets-with-Tokens-which-Carry-Data}
Ranko Lazic, Thomas~Christopher Newcomb, Jo{\"{e}}l Ouaknine, A.~W. Roscoe, and
  James Worrell.
\newblock Nets with tokens which carry data.
\newblock {\em Fundam. Inform.}, 88(3):251--274, 2008.

\bibitem{Forward-Analysis-for-Petri-Nets-with-Name-Creation}
Fernando Rosa{-}Velardo and David de~Frutos{-}Escrig.
\newblock Forward analysis for {P}etri nets with name creation.
\newblock In {\em Applications and Theory of {P}etri Nets, 31st International
  Conference, {PETRI} {NETS} 2010, Braga, Portugal, June 21-25, 2010.
  Proceedings}, pages 185--205, 2010.

\bibitem{What-Makes-Petri-Nets-Harder-to-Verify-Stack-or-Data}
Ranko Lazic and Patrick Totzke.
\newblock What makes {P}etri nets harder to verify: Stack or data?
\newblock In {\em Concurrency, Security, and Puzzles - Essays Dedicated to
  Andrew William Roscoe on the Occasion of His 60th Birthday}, pages 144--161,
  2017.

\bibitem{Coverability-Trees-for-Petri-Nets-with-Unordered-Data}
Piotr Hofman, Slawomir Lasota, Ranko Lazic, J{\'{e}}r{\^{o}}me Leroux, Sylvain
  Schmitz, and Patrick Totzke.
\newblock Coverability trees for {P}etri nets with unordered data.
\newblock In {\em Foundations of Software Science and Computation Structures -
  19th International Conference, {FOSSACS} 2016, Held as Part of the European
  Joint Conferences on Theory and Practice of Software, {ETAPS} 2016,
  Eindhoven, The Netherlands, April 2-8, 2016, Proceedings}, pages 445--461,
  2016.

\bibitem{DBLP:conf/lics/HofmanLT17}
Piotr Hofman, J{\'{e}}r{\^{o}}me Leroux, and Patrick Totzke.
\newblock Linear combinations of unordered data vectors.
\newblock In {\em 32nd Annual {ACM/IEEE} Symposium on Logic in Computer
  Science, {LICS} 2017, Reykjavik, Iceland, June 20-23, 2017}, pages 1--11,
  2017.

\bibitem{DBLP:conf/concur/HofmanL18}
Piotr Hofman and Slawomir Lasota.
\newblock Linear equations with ordered data.
\newblock In {\em 29th International Conference on Concurrency Theory, {CONCUR}
  2018, September 4-7, 2018, Beijing, China}, pages 24:1--24:17, 2018.

\bibitem{Silva1998}
Manuel Silva, Enrique Terue, and Jos{\'e}~Manuel Colom.
\newblock {\em Linear algebraic and linear programming techniques for the
  analysis of place/transition net systems}, pages 309--373.
\newblock Springer Berlin Heidelberg, Berlin, Heidelberg, 1998.

\bibitem{DBLP:conf/tacas/BlondinFHH16}
Michael Blondin, Alain Finkel, Christoph Haase, and Serge Haddad.
\newblock Approaching the coverability problem continuously.
\newblock In {\em Tools and Algorithms for the Construction and Analysis of
  Systems - 22nd International Conference, {TACAS} 2016, Held as Part of the
  European Joint Conferences on Theory and Practice of Software, {ETAPS} 2016,
  Eindhoven, The Netherlands, April 2-8, 2016, Proceedings}, pages 480--496,
  2016.

\bibitem{Petrinizer}
Javier Esparza, Rusl{\'a}n Ledesma-Garza, Rupak Majumdar, Philipp Meyer, and
  Filip Niksic.
\newblock An {SMT}-based approach to coverability analysis.
\newblock In Armin Biere and Roderick Bloem, editors, {\em Computer Aided
  Verification}, pages 603--619, Cham, 2014. Springer International Publishing.

\bibitem{DBLP:journals/tcs/Rosa-VelardoF11}
Fernando Rosa{-}Velardo and David de~Frutos{-}Escrig.
\newblock Decidability and complexity of {P}etri nets with unordered data.
\newblock {\em Theor. Comput. Sci.}, 412(34):4439--4451, 2011.

\bibitem{Karmarkar84}
Narendra Karmarkar.
\newblock A new polynomial-time algorithm for linear programming.
\newblock {\em Combinatorica}, 4(4):373--396, 1984.

\end{thebibliography}
 \section{Appendix}

\subsection{Proofs from section \ref{sec:Bound}}
\label{app:sec5}

\help*

\begin{proof}
By definition of $\uniformize$ operation, if $\alpha\not\in \setE$, then 
\begin{equation*}
\begin{split}
 \markf'(\bullet,\da) -\marki'(\bullet,\da)&= \sum_{j=0}^{|\setE|-1} \Delta(\rotating^{j}({\setE},\frac{\omega}{|\setE|}))(\bullet,\da)= \frac{c}{|\setE|} \left(\sum_{j=0}^{|\setE|-1}(\Delta(t)\cdot\Pmat)(\bullet,\da)\right)\\
&= c\Delta(t)\cdot \Pmat(\bullet, \da)= \markf(\bullet,\da)-\marki(\bullet,\da)
\end{split}
\end{equation*}
The second equality is due to the definition of $\rotate$, since data outside of $\setE$ are not touched by $\rotate$. Going further, if $\da\in\setE$ then $ \markf'(\bullet,\da) -\marki' (\bullet,\da)=$

\begin{equation*} \label{eq:uniform-diff}
\begin{split}
 \sum_{j=0}^{|\setE|-1} \Delta(\rotating^{j}({\setE},\frac{\omega}{|\setE|}))(\bullet,\da)=&
\frac{c}{|\setE|} \left(\sum_{j=0}^{|\setE|-1}(\Delta(t)\cdot\Pmat)(\bullet,next_{\setE}^j(\da))\right)\\
=\frac{c}{|\setE|} \left(\sum_{\db\in \setE}(\Delta(t)\cdot\Pmat)(\bullet,\db)\right)=&
\frac{\sum_{\db \in {\setE}}(\markf(\bullet,\db)-\marki(\bullet,\db))}{|\setE|}
\end{split}
\end{equation*}
\end{proof}
This completes the proof of Lemma~\ref{lem:help}. Next, we move to the important proof of Claim~\ref{claim:bnd} stated in Lemma~\ref{lem:decr}. We first recall the claim.

\claimClaimOne*
\begin{proof}[of Claim~\ref{claim:bnd}] We prove this claim by induction on $j$, the number of steps fired in the run $\sigma$. Assuming $\markm_j'$ to be as in the claim we show that  $\markm_{j+1}'$ satisfies the claim.

\noindent\textbf{Base Case} :- Initially (at marking $\markm_0=\marki=\markm_0'$). 
For all $\gamma\in\setE\cup\{\da\}$ $\marki(\bullet,\gamma)=\zero$. The two first points hold trivially and  for the third one we see that
$\zero=\frac{1}{|\setE|}\left(\sum_{\delta\in{\setE}\cup \{\da\}}\zero\right).$
Hence, shown.

\noindent\textbf{Induction step} :- Let us prove the three conditions in turn.
\begin{itemize}
    \item Condition 1. This is the simplest. Due to $\rotating(\da,\setE,\bullet)$ operations being a part of every step in $\rho$ we know that $\da$ does not participate in any step of $\rho$ so its value stays constant, and equals $\zero$. 

\item Condition 2. Due to the definitions of $\replace, \da,$ and $\setE$ we have that \linebreak   $\reducing(\da,\setE,\omega)(\bullet,\gamma)=\omega(\bullet,\gamma)$ holds for all $\gamma\in\dval(\marki)\cup\dval(\markf),$ where $\omega$ is any step in $\uniformyzing({\setE},\sigma_j)$.
Thus, if 
$\markm_j'\xrightarrow{\uniformyzing({\setE},\sigma_j)}_{\Q}\markz$, then we have:
\begin{equation}\label{eq:one}
  \markz(\bullet,\gamma)=\markm_{j+1}'(\bullet,\gamma)\text{ for all } \gamma\in\dval(\marki)\cup\dval(\markf).
\end{equation}

Now, by Lemma~\ref{lem:help} $\forall \gamma\in\dval(\marki)\cup \dval(\markf)$ we have $\markz(\bullet,\gamma)-\markm_j'(\bullet,\gamma)=\markm_{j+1}(\bullet,\gamma)-\markm_j(\bullet,\gamma)$. 
Further, by the induction hypothesis, we have \linebreak $\markm_j'(\bullet,\gamma)=\markm_j(\bullet,\gamma)$. Therefore $\markz(\bullet,\gamma)=\markm_{j+1}(\bullet,\gamma)$, and finally By Equation~(\ref{eq:one}) above, $\markm_{j+1}'(\bullet,\gamma)=\markm_{j+1}(\bullet,\gamma).$ 

\item Condition 3. This is the most complex condition to show. Let $\beta$ be the data value which is swapped with $\da$ in the $\replace$ operation, or any datum from $\setE$ such that $c_j\Delta(t_j)\cdot\Pmat_j)(\bullet,\da)=\zero.$
Suppose markings $\markz$ and $\markz'$ are such that $\markm_{j}'\xrightarrow{\sigma_j}\markz\text{ and 
}\markm_{j}'\xrightarrow{\reducing(\da,\setE,\sigma_j)}\markz'.$
Then, we observe that
\begin{enumerate}
    \item $\markz'(\bullet,\delta)=\markz(\bullet,\delta)$ for all $\delta\not\in\{\da,\beta\}$, 
    \item $\markz'(\bullet,\da)=\zero$ indeed from the induction assumption $\markm_j'(\bullet,\da)=\markm_j(\bullet,\da)$ $=\zero$ and according to definition $\replace$ the step $\reducing(\da,\setE,\sigma_j)$ does not touch tokens with data value $\da.$ 
    \item $\markz'(\bullet,\beta)=\markz(\bullet,\beta)+(\markz(\bullet,\da)-\markm_j'(\bullet,\da))=\markz(\bullet,\beta)+\markz(\bullet,\da),$ 
    the first equality is due to the definition of $\replace$ the second due to the induction assumption.  
\end{enumerate}

Now, combining above with Lemma~\ref{lem:help} for all $\gamma\in\setE$ we get,

\begin{equation*}
\begin{split}
\markm_{j+1}'(\bullet,\gamma)-\markm_j'(\bullet,\gamma) =\frac{1}{|\setE|}\left(\sum_{\delta \in {\setE}}\left(\markz'(\bullet,\delta)-\markm_j'(\bullet,\delta)\right)\right)=
\\
\frac{1}{|\setE|}\left(\markz(\bullet,\beta)+\markz(\bullet,\da)+\sum_{\delta \in \setE\setminus\{\da,\beta\}}\markz(\bullet,\delta)-\sum_{\delta\in\setE}\markm_j'(\bullet,\delta)
\right)  =
\\
\frac{1}{|\setE|}\left(
\sum_{\delta \in \setE\cup\{\da\}}
\markz(\bullet,\delta)-\sum_{\delta\in\setE}\markm_j'(\bullet,\delta)\right)\\
\end{split} 
\end{equation*}
Using $\markm_j(\bullet,\da)=\zero$ ( by induction assumption) and as $\markz-\markm_j'=$ $\markm_{j+1}-\markm_j$ we derive, $\markm_{j+1}'(\bullet,\gamma)-\markm_j'(\bullet,\gamma)= $
\begin{equation*}
\begin{split}
\frac{
\sum_{\delta \in \setE\cup\{\da\}}
\markz(\bullet,\delta)-
\sum_{\delta\in\setE\cup\{\da\}}\markm_j'(\bullet,\delta)}{|\setE|}
 =  \frac{
\sum_{\delta \in \setE\cup\{\da\}}
(\markm_{j+1}(\bullet,\delta)-\markm_j(\bullet,\delta))}{|\setE|}
.
\end{split} 
\end{equation*}
Moreover, for any $\gamma\in \setE$ we have 
$\markm_j'(\bullet,\gamma)=\frac{1}{|\setE|}\left(\sum_{\delta\in \setE\cup\{\da\}}\markm_j(\bullet,\delta)\right)$ by the induction hypothesis. Thus, we obtain $\markm_{j+1}'(\bullet,\gamma)-\markm_j'(\bullet,\gamma)=$
\begin{equation*}
\begin{split}
\frac{
\sum_{\delta \in \setE\cup\{\da\}}
\markm_{j+1}(\bullet,\delta)-\sum_{\delta\in\setE}\markm_j(\bullet,\delta)}{|\setE|}
=
\frac{
\sum_{\delta \in \setE\cup\{\da\}}
\markm_{j+1}(\bullet,\delta)
}{|\setE|}
-\markm_j'(\bullet,\gamma),
\end{split} 
\end{equation*}
from which we derive
$
\markm_{j+1}'(\bullet,\gamma)=
\frac{1}{|\setE|}\left(
\sum_{\delta \in \setE\cup\{\da\}}
\markm_{j+1}(\bullet,\delta)\right)$ as required.

\hfill(end of Proof of claim\ref{claim:bnd})\qed
\end{itemize}
\end{proof}

 \subsection{Proofs from section \ref{Sec:Q-reach}}
 \label{app:sec6}
\histadd*
 
 \begin{proof}
 We prove both properties separately.
 Let $\sum_{i=1}^{n}H_i = H_0$. We have \linebreak $\row(H_0)= \row(H_i)$ (the set of row indices for all $H_i$ is the same) and $\col(H_0) = \cup_{1\leq i\leq n} \col(H_i)$. Thus, for each $x \in \row(H_0)$ 
 $$\sum_{\alpha \in \col(H_0)} H_0(x,\alpha) = \sum_{i=1 }^{n} \sum_{\alpha \in \col(H_i)} H_i(x,\alpha) = \sum_{i=1}^{n} q_i.$$
 Hence, the first condition of Definition \ref{def:histogram} holds in $H$.
 
 Now we show that the second condition also holds finishing the proof. For each $\alpha \in \col(H_0)$, 
 $$\sum_{x \in \row(H_0)} H_0(x,\alpha) = \sum_{i=1}^{n} \sum_{x \in \row(H_0)}H_i(x,\alpha) \leq \sum_{i=1}^{n}q_i$$ 
 as each $H_i$ is a histogram. Thus the second property of the definition also holds and hence $H_0$ is a histogram with order $\sum_{i=1}^{n}q_i$. 

The proof of second property is very similar to the proof of Theorem 3 in \cite{DBLP:conf/lics/HofmanLT17}, the only difference is that our histograms have non-negative rational entries while there, histograms had natural entries. Here we only describe an overview of the complete argument. The proof relies on building a weighted bipartite graph from the histogram $H$ whose partite sets are row and column indices. An edge between nodes corresponding to row index $x$ and column index $\alpha$ is given a weight $H(x,\alpha)$. Now define a subset $D$ of $\col(H)$ as $D = \{\alpha ~| ~\sum_{x \in \row(H)} H(x,\alpha) = q \}$. Using Hall's theorem one can show that there exists matchings $M_1$, $M_2$ that saturate $\row(H)$, $D$ respectively. Using $M_1, M_2$ one can obtain another matching $M$ that saturates $\row(H) \cup D$. Now we take the minimum of edge weights in the matching, let that be $a_1$. Then we make two histograms from $H$ using $M$ and $a_1$ as follows. $M$ determines a set of row,column index pairs $E$ as follows: $(x,\alpha) \in E$ implies that the edge corresponding to the nodes representing $x$ and $\alpha$ is in $M$ and vice-verse. Construct a [$a_1$]-histogram $H_1$ having $\row(H_1) = \row(H), \col(H_1) = \col(H)$, and $H_1(x,\alpha) = a_1$ for $(x,\alpha) \in E$, $0$ otherwise. Modify $H$ by subtracting $a_1$ from all entries determined by $(x,\alpha) \in E$. Now we see that $H_1$ is an [$a_1$]-histogram and $H$ is a histogram of order $q-a_1$. We can repeatedly apply this procedure until $H$ becomes an [$a$]-histogram itself for some $a$. This completes the description of the proof.
\end{proof}

\QCharacterization*

\begin{proof}
Due to Theorem~\ref{thm:main2} if there is a run then there is run that uses at most $|\dval(\marki)\cup\dval(\markf)|+1+\max_{t \in T}(|\var(t)|)$ different data values. Due to Proposition~\ref{prop:histograms} there are required histograms. To prove the other direction, we just need to show that we can represent $\sum_{t \in T} \Delta(t) H_t $ as $\sum c_j \Delta(t_j) \Pmat_j$ having $c_j \in \Q^{+}$, as the latter is a sequence of $\Q$-steps and hence a $\Q$-run. To this end we just need to show that we can decompose a histogram as $H_i$ as $\sum_{k}c_k\Pmat_k$ for some constants $c_k \in \Q^{+}$ and some permutation matrices $\Pmat_k$. This follows from Lemma~\ref{lemma:hist add} as after decomposing $H_t$ into [$a_j$]-histograms we can take out $a_j$ and write $a_j \cdot \Pmat_j$ where $\Pmat_j$ is a permutation matrix. Thus we can decompose $H_t$ as $\sum_{k}c_k \Pmat_k$. This completes our proof. 
\end{proof}

\qreach*

\begin{proof}
We use the characterization from Lemma~\ref{lemma:Q-characterization}. We encode the reachability problem as a system of linear inequalities.
\begin{itemize}
\item $\markf-\marki = \sum_{t\in T}\Delta(t) H_t$.
\item Both the conditions of definition \ref{def:histogram} are to be satisfied for all the histograms $H_t$ , $t \in T$.
\item Variables are entries of the histograms and that is why for each variable $v$, we add condition $v \in \mathbb{Q}^+$. 
\end{itemize}
The total number of variables equals $|\setE| \cdot |T| \cdot |Var|$, since $|\setE|$ is polynomial (according to Lemma~\ref{lemma:Q-characterization}) the total number of unknowns is polynomial. Thus, the number of equations is also polynomial.
Since such a system of constraints can be solved as a system of linear equations over $\mathbb{Q}^{+}$ in Ptime in the size of input\cite{Karmarkar84}, the $\Q-$reachability can be solved in Ptime as the size of input is polynomial.
\end{proof}

 \subsection{Proofs from section \ref{Sec:CharacterizingQ+-reachability}}
 \label{app:sec71}
 
In this section, we prove Lemma~\ref{lemma:final charac}. We recall the statement now.

\lemmaFinalCharac*
The high level view of the proof is as follows. In the first step, we consider a special case when the $\Q$-reachability implies $\Q^+$-reachability between markings in Lemma~\ref{lemma:reachability_base_case} below.  The idea for this lemma and its proof is similar to Lemma 14 from \cite{DBLP:journals/fuin/FracaH15} (in fact it would be possible to make a reduction from our setting to the statement of the mentioned lemma but it would require restating definitions from \cite{DBLP:journals/fuin/FracaH15}). We extend it here for UDPN. The basic idea is to fire steps in such small fractions that the number of tokens never go negative. We repeatedly fire the complete $\Q$-run $\sigma$ with very small fractions until we reach the required marking. 
 The second step uses this lemma to show a weak characterization of $\Q^{+}$-reachability, without bounding the number of $\Q^+$-steps in Lemma \ref{lemma : first-way} below. Finally, the third step is to observe  that both $Pre(\sigma)$ and $Post(\sigma)$ can be bounded by  $P\times \dval(\sigma)$ we strengthen this result and obtain Lemma \ref{lemma:final charac}.

 \begin{restatable}{lemma}{reachabilityBaseCase}
 \label{lemma:reachability_base_case}
 If for $\Q^+$-markings $\marki$ and $\markf$, there exists a $\Q$-run $\sigma$ such that $\marki \xrightarrow{\sigma}_{\Q} \markf$ and $\forall (p,\da) \in Pre(\sigma), \marki(p,\da)>0$, $\forall (p,\da) \in Post(\sigma), \markf(p,\da) >0$, then $\markf$ is $\Q^+$-reachable from $\marki$.
 \end{restatable}
 \begin{proof}
For the $\Q$-run $\sigma= \{(c_i,t_i,P_i)\}_{|\sigma|}$, we define a constant  $\omega$, which is the sum of all tokens consumed and produced along the path $\sigma$:
\[\omega = \sum_{i=1}^{|\sigma|}\sum_{p \in P}{} \sum_{x \in \var{(t_i)}}  c_i\cdot( F(t_i,p)(x)+F(p,t_i)(x))\]
 Observe that for any factor $s\in \Q^+$ and any $\sigma'$ a prefix of $\sigma$ if $\markm\xrightarrow{s\cdot \sigma'}\markm'$ then $\markm'\geq \markm-s\cdot \omega;$ we use this inequality later in the proof.
Let a constant $c$ be a minimal distance from the empty marking to either $\marki$ or $\markf$, i.e. 
$$c = \min{\{\marki(p,\da), \markf(p,\db): (p,\da) \in Pre (\sigma), (p,\db) \in Post (\sigma)\}}.$$ 

 Let $n=\max(\lceil \frac{\omega}{c} \rceil,2).$
 Finally, we define the run the $\Q^+$-run $\rho$ by firing $n-$ times the run $\frac{1}{n}{\sigma}.$ 
 We claim that $\rho$ is the required $\Q^+$-run and is fireable at $\marki$. $\marki \xrightarrow{\rho}_{\Q^{}} \markf$ trivially holds. Hence, the only question that remains is its fire-ability. 
 
 To show $\Q^+$-fireability, 
 we consider intermediate markings:
 $$ \marki \xrightarrow{\frac{\sigma}{n}}_{\Q} \markm_1 \xrightarrow{\frac{\sigma}{n}}_{\Q} \markm_2 ... \markm_{n-1} \xrightarrow{\frac{\sigma}{n}}_{\Q} \markf .$$
First, we observe that each $\markm_i(p,\da)\geq c$  for every pair $(p,\da)\in Pre(\sigma)\cup Post(\sigma),$ indeed $\markm_i(p,\da)\geq min(\marki(p,\da),\markf(p,\da))\geq c$ for $(p,\da)\in Pre(\sigma)\cup Post(\sigma).$
So we only need to show fireability of the run $\frac{\sigma}{n}$ from $\markm_i$. But the number of tokens consumed along the run $\frac{\sigma}{n}$ is smaller than $\frac{\omega}{n}\leq \omega\cdot \frac{c}{\omega}=c$. So, the total number of consumed tokens along the run $\frac{\sigma}{n}$ is smaller than $c$ and smaller than the number of tokens in $\markm_i(p,\da).$ Thus $\frac{\omega}{n}$ is $\Q^+$ fireable.
\end{proof}
Now using the above Lemma, we show a weaker characterization of $\Q^{+}$-reachability, without bounding the number of $\Q^+$-steps. We formalize this as:

\begin{lemma} \label{lemma : first-way}
For two $\Q^{+}$-markings $\marki,\markf$, there exists a $\Q^{+}$-run $\sigma$ such that $\marki \xrightarrow{\sigma}_{\Q^{+}} \markf$ iff there exist markings $\marki'$ and $\markf'$ (possibly on a different run) such that 
\begin{enumerate}
\item $\marki'$ is $\Q^+$-reachable from $\marki$ 
\item There is a run $\sigma'$ such that $\dval(\sigma')\subseteq \dval(\sigma)$ and $\marki'\xrightarrow{\sigma'}_{\Q}\markf'.$
\item $\markf$ is $\Q^+$-reachable from $\markf'$ 
    \item $\forall (p,\da) \in Pre(\sigma'), \marki'(p,\da) > 0$ 
    \item $\forall (p,\da) \in Post(\sigma'), \markf'(p,\da) > 0$ 
\end{enumerate}
\end{lemma}

\begin{proof}
The easy direction is that the 5 conditions imply continuous reachability. Indeed, due to Lemma~\ref{lemma:reachability_base_case}, points $2,4,$ and $5$ imply continuous reachability from $\marki'$ to $\markf'$. Now, to obtain a fireable run from $\marki$ to $\markf$ we concatenate three runs: from $\marki$ to $\marki'$ (point 1), form $\marki'$ to $\markf'$, and the run from $\markf'$ to $\markf$ (point 3).


The proof in the opposite direction is more involved.
Before we start it, we introduce a new operation on two sequences $\{a_n\}, \{b_n\}$ where $\{a_n\}$ is a sequence of steps and $\{b_n\}$ is a sequence of real numbers, both having length $k$. We define  $a_n\combined b_n$ as $\{a_1\cdot b_1, a_2\cdot b_2,\ldots,
a_k\cdot b_k\}$.

    Let $\sigma = \{(c_i,t_i,P_i)\}_{|\sigma|}$ where terms have their usual meanings. 
    First, we need a small constant, namely a smallest positive number appearing in the problem definition divide by a biggest number in the problem definition 
    
    \begin{equation*}
        \begin{split}
            \omega = 
    \frac{\min(\{\marki(p,\da)>0, \markf(p,\da)>0:p\in P, \da\in \D\}\cup}{\max(\{\marki(p,\da)>0, \markf(p,\da)>0:p\in P, \da\in \D\}\cup}
    \\ 
    \frac{F(t,p)(x)>0, F(p,t)(x)>0:t\in T, x\in \var(t), p\in P\})}{\{F(t,p)(x)>0, F(p,t)(x)>0:t\in T, x\in \var(t), p\in P\})}.
        \end{split}
    \end{equation*}
    
    In addition we define $c_{min}=\frac{1}{2}\min(1, c_i:\text{ where } c_i \text{ are coefficients in } \sigma).$
    
    Let 
    $\setS_{prev}$ be a set of (place, datum) pairs which are consumed during the run but are not present in the initial configuration; similarly let $\setS_{post}$ be a set of (place, datum) pairs which are produced during the run but do not appear in the final configuration, i.e.
    $\setS_{prev} =   Pre(\sigma) \setminus \{(p,\da) : \marki(p,\da) > 0\}$ and 
    $\setS_{post} =   Post(\sigma) \setminus \{(p,\da) : \markf(p,\da) > 0\}.$
    For $(p,\da) \in \setS_{prev}$ 
     the value $\markm(p,\da)$ goes from 0 to non-negative value along the run. we call the first step after which $\markm(p,\da)$ becomes positive as the marking step, and similarly un-marking steps are the steps that make 
     $\markm(p,\da), ~(p,\da \in \setS_{post})$
    zero for the last time. We define $\sigma_{prev}$ and $\sigma_{post}$ two subsequences of $\sigma$. $\sigma_{prev}$ is the sequence of marking steps and $\sigma_{post}$ is the sequence of un-marking steps. 
    Finally, we define two other sequences $$\omega_{prev}=c_{min}\cdot\left(\frac{\omega}{2}\right), c_{min}\cdot\left(\frac{\omega}{2}\right)^2,  c_{min}\cdot\left(\frac{\omega}{2}\right)^3\ldots
     c_{min}\cdot\left(\frac{\omega}{2}\right)^{|\setS_{prev}|}\text{ and }$$
    $$\omega_{post}=c_{min}\cdot\left(\frac{\omega}{2}\right)^{|\setS_{post}|},c_{min}\cdot\left(\frac{\omega}{2}\right)^{|\setS_{post}|-1},c_{min}\cdot\left(\frac{\omega}{2}\right)^{|\setS_{post}|-2}\ldots c_{min}\cdot\left(\frac{\omega}{2}\right).$$

    \begin{myclaim}
    $\sigma_{prev}\combined \omega_{prev}$ can be fired from $\marki$.
    \end{myclaim}
   \begin{myclaim}
    $\markm$ obtained after firing $\sigma_{prev}\combined \omega_{prev}$ from $\marki$ is positive on all elements in $Pre(\sigma).$
   \end{myclaim}
    
    Indeed, the coefficients provided by the $\omega_{prev}$ sequence guarantee that any place that was marked during the run $\sigma_{prev}\combined \omega_{prev}$
    will nether get negative nor zero during the run $\sigma_{prev}$. The constant $\omega$ is used to reduce difference between the minimal amount of tokens that can be produced and the maximal amount of tokens that can be consumed in consecutive steps.
    
    Similarly, we claim that there is $\markm'$ such that $\sigma_{post}\combined \omega_{post}$ can be fired firm $\markm'$ and it leads to $\markf$. Indeed, it suffices to reverse direction of all transitions and look to $\sigma_{post}\combined \omega_{post}$ backward.
    
    Moreover, there is a $\Q$ run $\delta$ form $\markm$ to $\markm'$ such that 
    $\dval(\delta)=\dval(\sigma)$. 
    Indeed,
    $\delta$ can be obtained via removing from $\sigma$ two sequences $\sigma_{post}\combined \omega_{post}$ and $\sigma_{prev}\combined \omega_{prev}$. It is possible as due to constant $c_{min}$ the coefficients of any step in runs $\sigma_{post}\combined \omega_{post}$ and $\sigma_{prev}\combined \omega_{prev}$ are smaller than coefficients in the run $\frac{1}{2}\sigma.$ 
    Finally, we put $\marki'=\markm$,
    $\markf'=\markm',$ and $\sigma'=\delta$ as $\markm,\markm'$ and $\delta$ satisfies assumptions of Lemma~\ref{lemma:reachability_base_case}. This completes the proof of Lemma~\ref{lemma : first-way}.
\end{proof}

Now to obtain the proof Lemma \ref{lemma:final charac} from the above Lemma, 
we analyze lengths of $\sigma_{prev}$
and $\sigma_{post}$ in the proof of the above Lemma. Trivially, both lengths are bounded by $|Pre(\sigma)|$ and $|Post(\sigma)|$, respectively, as each step
introduces or removes a new pair $(p,\da)$.
Further, both $|Pre(\sigma)|$ and $|Post(\sigma)|$ can be bounded by 
$|P\times \dval(\sigma)|$ which is exactly what we require in the formulation of Lemma~\ref{lemma:final charac}.
 
\subsection{Proofs from Section \ref{section:trans7.2}}
\label{section:transformation}

We show that every UDPN can be converted to a \loopless\ UDPN with the required equivalence.
\textbf{Transformation to a \loopless\ UDPN}
\begin{restatable}{lemma}{transformation}\label{lemma:transformation}
For every net $N$ and two markings 
$\marki,\markf$
in polynomial time one can construct a \loopless\ net $N'$ and two markings $\marki',\ \markf'$ such that
$\marki\xrightarrow{}_{\Q^+}\markf$ in the net $N$ iff $\marki'\xrightarrow{}_{\Q^+}\markf'$ in $N'.$
\end{restatable}

\begin{proof}
We first construct the \loopless\ net and then show its equivalence.\\
Let the initial net be $N = \Net{P,T,F,\Var}$ and markings be $\marki,\markf$ and transformed net $N' = (P_c,T_c,F_c, \Var)$ and markings be $\marki',\markf'$. The construction is as follows.
\begin{enumerate}
    \item $P_c = P \cup P'$ where $|P'|=|P|$, and for each place $p \in P$  there is a corresponding place denoted as $f(p)$, where  $f$ is a relabelling operation.  $P'$ is defined as   $P'= \cup_{p \in P} f(p)$.
    Note that $|P_c|=2 \cdot |P|$.
    \item $T_c$ contains a modified transition corresponding to $T$ and an additional transition for each place. We add a transition for each place in $t$ that can remove an any data token from $f(p)$ and add it to $p$. We modify each transition $t\in T$ so that if a place $p \in PrePlace(t)\cap PostPlace(t)$ , we remove $p$ from the PostPlace and add $f(p)$ to it. This is reflected in flow relation $F_c$- if a place $p \in PrePlace(t)\cap PostPlace(t)$ , $F_c(t,f(p))=F(t,p)$ and  $F_c(t,p)=\emptyset$.
    Otherwise, $F_c(t,p)=F(t,p)$ and $F_c(p,t)=F(p,t)$. Further we add $|P|$ transitions. For each $p \in P$ we define a transition as $t$ having pre-place as $f(p)$ and post-place as $p$, we add the relation $(f(p),t) \to ({x}\to 1), x\in Var$ and $(t,p) \to (x\to 1)$ in $F_c$. This completes the construction for $T_c,F_c$. Note that $|T_c|=|T|+|P|$.
    \item We define $\marki'$ as in $\marki$ for all $p \in P \cap P_c$ and for the $\forall p \in P_c\backslash P$, we define the marking to have zero tokens for all data.i.e. $\forall p \in P,\forall d \in \D, \marki'(p,d)=\marki(p,d)$ and  $\forall p \in P_c\backslash P, \forall d \in \D,  \marki'(p,d)=0 $. Similarly we define $\markf'$.
\end{enumerate}
\begin{myclaim}
The $\Q^+$-reachability problem on $N$ and $N'$  is equivalent.
\end{myclaim}  
\begin{proof}
Suppose that in the net $N$, $\markf$ is $\Q^+$-reachable. Then we make the following modifications to the $\Q^+$-run :- We fire a transition $t$ as in the original run, after which we fire all the newly added transitions $t'$ which are involved with only $p$ and $f(p)$) with appropriate modes so that for all $ f(p)$ all the data tokens are removed from $f(p)$ and added to $p$. This is possible due to the fact that the flow relation for all such $t'$ has only one variable in both arcs. We repeat this step for all transitions in the run.  With this modification, each marking in the run has exactly the same tokens $\forall p \in P$  as in the original run after firing the transitions and $0$ $\forall p \in P_c\backslash P$. Since $f(p)$ is not a pre-place for any transition $t$ in the original net, all transitions can be fired.  By induction, we reach a $\markf'$ having the above mentioned property corresponding to final marking $\markf$. The marking $\markf'$ is as described by the transformation.  Therefore, $\markf'$ is $\Q^+$-reachable in $N'$.\\
In the other direction, suppose it is $\Q^+$-reachable in $N'$, then whenever a new transition $t$ is fired, a modified transition $t_1$ must have been fired previously. Therefore, we remove all the firings of new transitions (take a projection on $T$), and show that it remains a valid run. For a new transition to have been fired, a transition must have been fired that must have put tokens in the new place. However, in the original net, the tokens were simply added in the old place. Therefore, the transition can still be fired. Hence shown.
\end{proof}
This completes the proof of lemma \ref{lemma:transformation}.
\end{proof}

\begin{example}
Consider the net $N$ in Figure~\ref{fig:loopless} (left). Then the net we get after the transformation is $N'$ in Figure~\ref{fig:loopless} (right).

\begin{figure}[t]
\begin{center}
{\begin{tikzpicture}[node distance=1.3cm,>=stealth',bend angle=45,auto]
\centering
  \tikzstyle{place}=[circle,thick,draw=blue!75,fill=blue!20,minimum size=6mm]
  \tikzstyle{red place}=[place,draw=red!75,fill=red!20]
  \tikzstyle{transition}=[rectangle,thick,draw=black!75,
  			  fill=black!20,minimum size=4mm]

  \begin{scope}
    \node [place,label=$p_2$] (w1)[xshift=0mm,yshift=0mm]{};
     \node [transition] (t1) [left of=w1,,xshift=0mm,yshift=-10mm] {t}
      edge [pre] node{x} (w1)
      edge [post,bend right] node{z} (w1);
    \node [place,label=$p_1$] (w2)[left of=t1,xshift=0mm,yshift=10mm]{}
    edge [pre, bend right] node{x}(t1)
    edge [post] node{y}(t1);
    \node [place] (p3)[below of=w1,xshift=0mm,yshift=-10mm]{}
    
    edge [pre] node{x,z}(t1);
   
      \node [place,label=$p_3$] (p4)[below of=w2,xshift=0mm,yshift=-10mm]{}
      edge [pre] node{$\{2y\}$}(t1);
    
  \end{scope}
\end{tikzpicture}
}
\scalebox{0.83}{
\begin{tikzpicture}[node distance=1.3cm,>=stealth',bend angle=45,auto]
\centering
  \tikzstyle{place}=[circle,thick,draw=blue!75,fill=blue!20,minimum size=6mm]
  \tikzstyle{red place}=[place,draw=red!75,fill=red!20]
  \tikzstyle{transition}=[rectangle,thick,draw=black!75,
  			  fill=black!20,minimum size=4mm]

  \begin{scope}
    \node [place,label=$p_2$] (w1)[xshift=0mm,yshift=0mm]{};
     \node [transition] (t1) [left of=w1,,xshift=0mm,yshift=-10mm] {t}
      edge [pre] node{x} (w1);
    \node [place,label=$p_1$] (w2)[left of=t1,xshift=0mm,yshift=10mm]{}
    edge [post] node{y}(t1);
    \node [place] (p3)[below of=w1,xshift=0mm,yshift=-10mm]{}
    
    edge [pre] node{x,z}(t1);
   
      \node [place,label=$p_3$] (p4)[below of=w2,xshift=0mm,yshift=-10mm]{}
      edge [pre] node{$\{2y\}$}(t1);
      
      \node [place,label=$f(p_1)$] (w5)[left of=t1,xshift=-10mm,yshift=0mm]{}
      edge [pre] node{x} (t1);
      \node [place,label=$f(p_2)$] (w6)[right of=t1,xshift=10mm,yshift=0mm]{}
      edge [pre] node{z} (t1);
      \node [place,label=$f(p_3)$] (w7)[left of=p4,xshift=-20mm,yshift=0mm]{};
      \node [place,label=$f(p_4)$] (w8)[right of=p3,xshift=20mm,yshift=0mm]{};
      \node [transition] (t2)[right of=w7,xshift=0mm,yshift=0mm]{}
      edge [pre] node{x}(w7)
      edge [post] node{x}(p4);
      \node [transition] (t3)[right of=p3,xshift=0mm,yshift=0mm]{}
      edge [pre] node{x}(w8)
      edge [post] node{x}(p3);
      \node [transition] (t4)[right of=w1,xshift=10mm,yshift=0mm]{}
      edge [pre] node{x}(w6)
      edge [post] node{x}(w1);
      \node [transition] (t5)[left of=w2,xshift=-10mm,yshift=0mm]{}
      edge [pre] node{x}(w5)
      edge [post] node{x}(w2);
    
  \end{scope}
\end{tikzpicture}
}
\end{center}
\caption{a UDPN $N$ (left) and its transformed net $N'$ (right)}
\label{fig:loopless}
\end{figure}
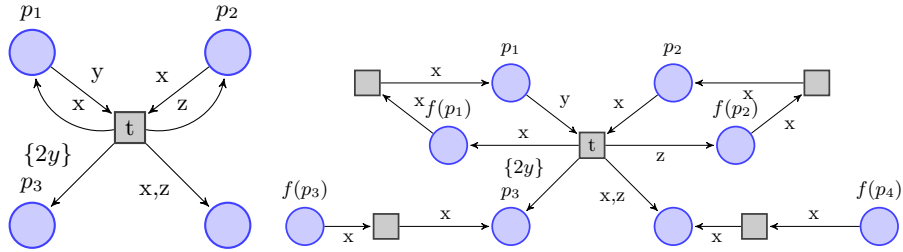
\end{example}

\Qstep*

\begin{proof}
Recall that by lemma \ref{lemma:hist add}, every histogram $H$ can be decomposed as $H=\sum c_i\Pmat_i$. Therefore, applying this decomposition we get $\markf-\marki=\Delta(t) \cdot \sum c_i \Pmat_i$. Consider a $\Q^{+}$-run $\sigma=\run{c_i, t, \Pmat_i}{|\sigma|}$ from $\marki$ to $\markf$. We want to show that $\marki \xrightarrow{\sigma}_{\Q^+} \markf$ holds.
As the net is \loopless\ we can split places into three kinds: places from which the run $\sigma$ consumes tokens, to which the run $\sigma$ produces, and places not touched by the run $\sigma$.
For the second and third kind of places it is trivial that number of tokens with any data value is not getting negative along $\sigma.$ 
For the place of the first kind and for any data value observe that, along the run $\sigma$ the number of tokens in the place and with the datum can only drop. Thus, if at any moment along the run $\sigma$ it got negative then it would stay negative to the very end of $\sigma$. But in the end i.e. $\markf$ it is non-negative.
Thus, the number of tokens with any data value in any place along $\sigma$ stays non-negative, and $\marki \xrightarrow{\sigma}_{\Q^+}  \markf$ holds.
\end{proof}

\end{document}